\documentclass[12pt]{article}
\usepackage{amsmath,amssymb,amsthm,amsxtra,overpic,bbm,bm,epsfig,subfigure}
\usepackage{hyperref}
\usepackage{color}
\usepackage{stackengine} 
\usepackage{multirow}
\usepackage{slashed}

\begin{document}
\vspace{0.2cm}

\begin{center}
{\Large\bf Towards a systematic study of non-thermal leptogenesis from inflaton decays}\\
\vspace{0.2cm}
\end{center}
\vspace{0.2cm}

\begin{center}
{\bf Xinyi Zhang}~$^{a,b,c,d}$~\footnote{Email: zhangxy@hbu.edu.cn}
\\
\vspace{0.2cm}
$^a$ {\em \small Department of Physics, Hebei University, Baoding, 071002, China}\\
$^b$ {\em \small Hebei Key Laboratory of High-precision Computation and Application of Quantum Field Theory, Baoding, 071002, China}\\
$^c$ {\em \small Hebei Research Center of the Basic Discipline for Computational Physics, Baoding, 071002, China}\\
$^d$ {\em \small Institute of High Energy Physics, Chinese Academy of Sciences, Beijing 100049, China}
\end{center}

\vspace{1.5cm} 
 
\begin{abstract}
This paper investigates non-thermal leptogenesis from inflaton decays in the minimal extension of the canonical type-I seesaw model, where a complex singlet scalar $\phi$ is introduced to generate the Majorana masses of right-handed neutrinos (RHNs) and to play the role of inflaton. First, we systematically study non-thermal leptogenesis with the least model dependence. We give a general classification of the parameter space and find four characteristic limits by carefully examining the interplay between inflaton decay into RHNs and the decay of RHNs into the standard-model particles. Three of the four limits are truly non-thermal, with a final efficiency larger than that of thermal leptogenesis. Two analytic estimates for these three limits are provided with working conditions to examine the validity. In particular, we find that the {\it strongly non-thermal RHNs} scenario occupies a large parameter space, including the oscillation-preferred $K$ range, and works well for a relatively-low reheating temperature $T^{}_{\rm RH} \geq 10^3~{\rm GeV}$, extending the lower bound on the RHN mass to $2\times 10^{7}~{\rm GeV}$. The lepton flavor effects are discussed. 
Second, we demonstrate that such a unified picture for inflation, neutrino masses, and baryon number asymmetry can be realized by either a Coleman-Weinberg potential (for the real part of $\phi$) or a natural inflation potential (for the imaginary part of $\phi$). The allowed parameter ranges for successful inflation and non-thermal leptogenesis are much more constrained than those without inflationary observations. We find that non-thermal leptogenesis from inflaton decay offers a testable framework for the early Universe. It can be further tested with upcoming cosmological and neutrino data. The model-independent investigation of non-thermal leptogenesis should be useful in exploring this direction.
\end{abstract}
 
\newpage 
 
\section{Introduction}
  
The Standard Model of particle physics (SM) provides the best description of particle interactions and has been successfully tested at an impressive accuracy. There are several reasons to believe that the SM is only an effective field theory at the electroweak scale. The mass of the neutrino is solid evidence. The neutrino oscillation phenomenon indicates that neutrinos have a tiny mass. The two squared-mass differences are measured to be~\cite{Esteban:2020cvm}
\begin{align}
\Delta m_{21}^2=\left( 7.42^{+0.21}_{-0.20}\right)\times 10^{-5}~\mathrm{eV}^2,\quad \Delta m_{31}^2=\left( 2.515\pm 0.028\right)\times 10^{-3}~\mathrm{eV}^2 \label{eq:neutrino_obs}
\end{align}
for the normal ordering of neutrinos.
The tiny neutrino mass can be nicely explained through the seesaw mechanism. In the simplest type-I seesaw, at least two heavy right-handed neutrinos (RHNs) are introduced and have a Yukawa coupling to the light neutrinos. After the RHNs are integrated out, the light neutrinos get a small mass. Interestingly, including the RHNs can help to solve another beyond SM question: the origin of the baryon asymmetry of the Universe (BAU).

The fact that the Universe consists only of matter and the consideration that the Universe should start with an equal amount of matter and anti-matter trigger the question of the origin of BAU. The current value of the baryon asymmetry precisely measured in CMB can be expressed in terms of the baryon-to-entropy ratio as~\cite{Planck:2018vyg}
\begin{align}
Y_\mathrm{B}^{}=\frac{n_\mathrm{B}^{}}{s}=\left(8.72\pm 0.08\right)\times 10^{-11}\;,\label{eq:BAU_obs}
\end{align}
where $n_\mathrm{B}^{}$ is the baryon number density and $s$ is the entropy density.
Leptogenesis is one of the mechanisms generating the BAU dynamically (for reviews, see, e.g., Refs.~\cite{Giudice:2003jh,Buchmuller:2004nz,Davidson:2008bu,Xing:2020ald,Bodeker:2020ghk,Xing:2011zza}).
In leptogenesis~\cite{Fukugita:1986hr}, the CP and lepton-number-violating decay of the RHNs generates a lepton asymmetry, which is converted to a baryon asymmetry by the electroweak sphaleron processes. Leptogenesis works closely with neutrino physics. On the one hand, the heavy RHNs help to understand the small active neutrino mass through the type-I seesaw mechanism. On the other hand, the CP-violating phases as probed in neutrino oscillation or neutrinoless double beta decay experiments can be the source of CP violation needed in leptogenesis~\cite{Branco:2001pq,Branco:2002xf,Pascoli:2006ie,Pascoli:2006ci,Moffat:2018smo,Xing:2020erm,Xing:2020ghj,Zhang:2020lir}. Consequently, leptogenesis has been investigated in many neutrino models (for a few examples, see, e.g., Refs.~\cite{Bi:2003yr,Xing:2006ms,Guo:2006qa,Gu:2010xc,Gehrlein:2015dxa,Gehrlein:2015dza,Sarma:2022qka,Mahapatra:2023dbr}).
Most literature focuses on thermal leptogenesis, which refers to the cases in which the RHNs are in or produced in thermal equilibrium first, then decay out of equilibrium. In the former case, the RHNs have an abundance following the thermal distribution, while in the latter case, a zero one such that they are produced in the thermal bath by scattering with other SM particles. The RHNs can be produced non-thermally, i.e., the so-called non-thermal leptogenesis. The idea of non-thermal leptogenesis was first proposed in Ref.~\cite{Lazarides:1990huy} following inflaton decay.

There is growing evidence that the early Universe undergoes an exponentially-accelerated expansion era, i.e., inflation, which solves the flatness and the horizon problem~\cite{Guth:1980zm}. More importantly, the quantum fluctuations during inflation seed the anisotropy as observed in cosmic microwave background (CMB) and the baryon acoustic oscillations (BAO) measurements (for reviews, see, e.g., Refs.~\cite{Bassett:2005xm,Baumann:2009ds}). There is a consensus that inflation did happen, but a standard inflation model is still missing. In the simplest realization of inflation, i.e., the single-field slow-roll models, the accelerated expansion happens when the Universe is dominated by the potential energy of a scalar field, i.e., the inflaton, which rolls slowly towards its potential minimum. A few parameters usually characterize the dynamics of these models. Consequently, these models are being actively tested with the CMB and the BAO data, and some of them have been ruled out. 

Non-thermal leptogenesis naturally (but not necessarily) connects to inflation. There are many studies connecting inflation and non-thermal leptogenesis. For example, non-thermal leptogenesis is discussed in supersymmetric hybrid inflation models~\cite{Kumekawa:1994gx,Asaka:1999yd,Lazarides:1999dm,Jeannerot:2001qu,Senoguz:2003hc,Dent:2003dn,Dent:2005gx,Endo:2006nj,Antusch:2010mv,Khalil:2012nd}, chaotic inflation~\cite{Murayama:1992ua,Ellis:2003sq,Pallis:2011ps,Pallis:2011gr,Pallis:2012iw}, hilltop inflation~\cite{Antusch:2018zvu}, Coleman-Weinberg potential~\cite{Panotopoulos:2021ttt,SravanKumar:2018tgk}, supersymmetric  SO($10$) models~\cite{Fukuyama:2005us,Baer:2008eq,Fukuyama:2010hh}, and various neutrino models~\cite{Asaka:2002zu,Allahverdi:2002gz,Panotopoulos:2006wj,Senoguz:2007hu}. Ref.~\cite{Ghoshal:2022fud} investigates constraints from non-thermal leptogenesis on inflationary observables. Generic studies can be found in Refs.~\cite{Chung:1998rq,Hahn-Woernle:2008tsk}. Ref.~\cite{Chung:1998rq} discusses non-thermal leptogenesis compared to thermal and preheating cases and provides analytical estimates. Ref.~\cite{Hahn-Woernle:2008tsk} focuses on the case that $M_1^{} \sim T_\mathrm{RH}^{}$ where $M_1^{}$ is the lightest RHN mass and $T_\mathrm{RH}^{}$ is the reheating temperature.

Given the situation that the inflationary models are being tested and we also have the BAU and neutrino data, we consider one simple but intriguing possibility that inflaton couples to RHNs and the RHNs decay reheats the Universe through their subsequent decays to SM particles. Non-thermal generation of RHNs is crucial to connect the inflation observational constraints to BAU and oscillation data. This work presents the first systematic study of the whole parameter space in non-thermal leptogenesis. We give a general classification of the parameter space and find four characteristic limits. The dynamics of the limits are examined both analytically and numerically. We identify the working conditions for analytic estimates and check the intermediate scenarios numerically. Flavor effects in non-thermal leptogenesis are considered for the first time. We find the previously overlooked {\it ``strongly non-thermal RHNs"} limit is quite relevant in non-thermal leptogenesis. We give two examples of inflation models fulfilling the connection. In both models, we find viable parameter space satisfying the constraints in Eq.~(\ref{eq:neutrino_obs}) and Eq.~(\ref{eq:BAU_obs}), as well as inflationary constraints, indicating that such a connection is valid and worthy of further study.

There are several reasons to motivate such a study. First, the ingredients (inflation, type-I seesaw, leptogenesis) are well-motivated physics themselves, offering simple and elegant solutions to important questions in cosmology and particle physics. Second, the connection appears natural and economical. The same physics can be addressed without introducing other degrees of freedom. Third, the connections offer additional constraints compared with treating them separately. Only non-thermal leptogenesis with the memory of initial conditions allows such a combined constraint. With growing cosmological and neutrino data, we can test this path toward the early Universe, completing our understanding of the physics of the Universe.

This paper is organized as follows. In Sec.~\ref{sec:nt}, we perform a systematic investigation of the non-thermal leptogenesis dynamics on a general basis. Analytical estimates of the final baryon asymmetry are given in Sec.~\ref{sec:analytic}, where four characteristic scenarios with conditions to distinguish them are proposed. A numerical investigation of the Boltzmann equations is given in Sec.~\ref{sec:numeric}. Following that, we discuss the final efficiency (Sec.~\ref{sec:efficiency}), the lower limit of the RHN mass (Sec.~\ref{sec:numass}), and the flavor effect (Sec.~\ref{sec:flavor}) and then give a summary in Sec.~\ref{sec:sum}. In Sec.~\ref{sec:inflation}, we give the results when concrete inflation models and observations are included. In Sec.~\ref{sec:cwi}, we consider the inflaton to be the real part of $\sigma$ and acquire a Coleman-Weinberg potential. In Sec.~\ref{sec:ni}, the inflaton is the imaginary part of the $\sigma$, i.e., the Majoron, and has a natural inflation potential. We conclude in Sec.~\ref{sec:conclusion}.
 
\section{Non-thermal leptogenesis}\label{sec:nt}
In this section, we perform a general study on non-thermal leptogenesis from inflaton decay without identifying inflaton. We assume that the inflaton couples only directly to the RHNs. The following discussion can be readily adapted to the cases where the inflaton decays to the RHNs with a branching ratio $B_N^{}$. Our discussion corresponds to $B_N^{}=1$. 

After inflation, inflaton oscillates coherently around its potential minimum. As it couples directly to the RHNs, once its decay rate $\Gamma_\phi^{}$ exceeds the Universe's expansion rate, it will decay effectively into the RHNs. 
Eventually, the RHNs will undergo CP and lepton-number-violating decays into charged leptons and Higgs, which generate a non-zero lepton asymmetry converting to a non-zero baryon asymmetry when electroweak sphaleron processes are effective. We are interested in clarifying the processes between inflaton decay and the baryon asymmetry generation. Our main objective in this section is to answer the following two questions: (i) What characteristic evolutionary processes can we have? (ii) What are the most relevant quantities characterizing the processes? The answers to these questions are not straightforward. For example, one may expect an RHN matter-dominated era; does it happen for both strong and weak Yukawa interactions? The answer to the second one is even more tricky.

Without loss of generality, we assume that the RHNs are hierarchical, and thus only the lightest one with mass $M_1^{}$ is relevant for leptogenesis. Extending to other possible RHN mass spectra will be similar to that in thermal leptogenesis. 
In the following, we perform a systematic study and then present our answers to these two questions in Sec.~\ref{sec:sum}.


\subsection{Relevant quantities}
We start with introducing relevant quantities. The inflaton decay rate is
\begin{align}
\Gamma_\phi^{} = y_N^2 \frac{M_\phi^{}}{4\pi} \;,\label{eq:Gamma_phi}
\end{align}
where $y_N^{}$ is the coupling of inflaton $\phi$ with RHN and $M_\phi^{}$ is the inflaton mass.
Defining the reheating temperature $T_\mathrm{RH}^{}$ to be the temperature at the beginning of the radiation-dominated era and assuming that the RHNs decay instantly\footnote{This assumption is to assure consistency. In case the RHNs decay for some time, a more accurate treatment should define $H(T_\mathrm{RH}^{})=\Gamma_N^{}$ with $\Gamma_N^{}$ being the RHN decay rate.}, $T_\mathrm{RH}^{}$ can be calculated from $H(T_\mathrm{RH}^{})=\Gamma_\phi^{}$ as
\begin{align}
T_\mathrm{RH}^{} = \displaystyle \left(\frac{45}{4\pi^3 g_*^{}} \right)^{\frac{1}{4}} \sqrt{\Gamma_\phi^{} m_\mathrm{pl}^{}}\;, \label{eq:Trh}
\end{align}
where $g_*^{}$ is the effective degrees of freedom in the thermal radiation bath and $m_\mathrm{pl}^{}$ is the Planck mass. We use the Hubble parameter in the radiation-dominated era $H =  \sqrt{4\pi^3g_*^{}/45} T^2/m_\mathrm{pl}^{}$. Notice that during reheating, there might have a higher temperature than this~\cite{Chung:1998rq}, but during that period, the Universe is not radiation dominated. If there is an RHN or other matter-dominated period after inflation, but before radiation domination, the reheating temperature should be defined according to that decay rate. For definiteness, we will always use $T_\mathrm{RH}^{}$ as defined by the inflaton decay rate. This definition relates $\Gamma_\phi^{}$ and $T_\mathrm{RH}^{}$, and it allows us to parameterize the coupling $y_N^{}$ in terms of $T_\mathrm{RH}^{}$. 

The RHN decay rate in its rest frame reads
\begin{align}
\widetilde{\Gamma}_{N}^{} =\displaystyle \frac{\left(Y_\nu^\dagger Y_\nu^{} \right)_{11}^{}}{8\pi } M_1^{}\;.\label{eq:GammaNt}
\end{align}
When leptogenesis is concerned, one should include a thermally-averaged Lorentz dilation factor which takes into account the RHN distribution and enables a $z \equiv M/T$ dependence, 
\begin{align}
\Gamma_N^{}=\displaystyle\widetilde{\Gamma}_{N}^{} \frac{K_1^{}(z)}{K_2^{}(z)} \;,\label{eq:GammaNNt}
\end{align}
where $K_1^{}(z)/K_2^{}(z)$ is the thermally-averaged dilation factor with $K_1^{},~K_2^{}$ being modified Bessel functions and $z=M_1^{}/T$. It is useful to parameterize the decay rate as~\cite{Hahn-Woernle:2008tsk}
\begin{align}
\Gamma_N^{}=H(M_1^{}) K \frac{K_1^{}(z)}{K_2^{}(z)}\;, \label{eq:GammaN}
\end{align}
where $K$ is the decay parameter and is defined as $K=\widetilde{\Gamma}_N^{}/H(M_1^{})=\left(Y_\nu^\dagger Y_\nu^{} \right)_{11}^{} M_1^{}/\left[8\pi H(M_1^{}) \right]$ \footnote{In literature, $K$ is also defined as the ratio of two quantities with mass dimension and comparable to the light neutrino mass, i.e., 
$K=\widetilde{m}_1^{}/m_*^{}$ with
\begin{align}
\widetilde{m}_1^{} &= \displaystyle \frac{\left(Y_\nu^\dagger Y_\nu^{} \right)_{11}v^2}{M_1} =\frac{8\pi v^2}{M_1^2} \widetilde{\Gamma}_{N}^{} \;, \label{eq:mtilde}\\
m_*^{} &= \displaystyle \frac{8\pi v^2}{M_1^2} H(M_1^{}) \;, \label{eq:mstar}
\end{align}
where $v$ is the vacuum expectation value (vev) of the SM Higgs.},
which relates closely to the Yukawa couplings and effectively characterizes the strength of the wash-out processes, as will be specified later.
Let $\widetilde{\Gamma}_{N}^{} = H(T_*^{})$, we introduce the decay temperature
\begin{align} 
T_*^{} = \sqrt{K} M_1^{}\;.\label{eq:Tstar}
\end{align}

The final baryon asymmetry originating from the RHNs decay can be expressed as 
\begin{align}
Y_\mathrm{B}^{}=\displaystyle \frac{n_\mathrm{B}^{}}{s} =\frac{c_\mathrm{sph}^{} \epsilon}{s} n_N^{} \;,
\end{align}
where $s=g_*(2\pi^2/45) T^3$ is the entropy density, $c_\mathrm{sph}^{}=28/79$ is the B$-$L to $\mathrm{B}$ converting factor in sphaleron processes with $\mathrm{B}$ being the baryon number and $\mathrm{L}$ being the lepton number, $\epsilon$ is the CP asymmetry generated by RHNs decay, it quantifies the lepton number generated per RHN decay, $n_\mathrm{B}^{}$ is the baryon number density and $n_N^{}$ is the RHN number density that converts to a net $\mathrm{L}$ number. The main task is determining $n_N^{}$ that generates a net $\mathrm{L}$ number.

Our considered system includes $\{\phi, N, R \}$ where $R$ denotes the radiation generated by RHNs decay. The evolution of this system, and consequently, $n_N^{}$ we are interested in, is dictated by a set of Boltzmann equations. Analytical estimations apply to some limits, providing useful insights into physics. In what follows, we first work in the analytic limits and then resort to the Boltzmann equations.

\subsection{Analytical estimations and classifications of the scenarios}\label{sec:analytic}

We consider first the $K\ll 1$ limit, which is the typical non-thermal limit since the Yukawas are so weak that the Yukawa interactions cannot thermalize the RHNs (we neglect other interactions). Consequently, the generated lepton number asymmetry from the RHN decay all survives, and we can infer $n_N^{}$ without resorting to the Boltzmann equation system. The key task for this limit is to determine the time (temperature) when the RHNs actually decay.

For inflaton decays to RHNs to be kinetically allowed, $M_\phi^{} \geq 2 M_1^{}$ should be satisfied. Depending on the relative magnitudes of $M_\phi^{}$ and $M_1^{}$, the RHNs can be produced relativistically or non-relativistically.
If there is a hierarchy of the inflaton mass and the RHN mass, i.e., $M_\phi^{}\gg M_1^{}$, the RHNs are produced relativistically, with energy $E_N^{} \simeq M_\phi/2$. As the Universe cools down, the RHNs become non-relativistic at a temperature 
\begin{align}
T_\mathrm{NR}^{}=T_\mathrm{RH}^{} M_1^{}/E_N^{}, \label{eq:Tnr}
\end{align}
and dominate the energy density of the Universe as matter.\footnote{It only happens for the RHNs are the only decay product of the inflation. In a more general setup, where the inflaton also decays to other particles (presumably as radiation), the RHNs coexist with the radiation bath before they decay. When the RHNs become non-relativistic in this scenario, their energy density will redshift like matter and, eventually, dominate the energy density of the Universe if they did not decay before.} We call it a ``RHN-dominated" case.

The RHNs decay at the decay temperature when $T_*^{} < T_\mathrm{NR}^{} < T_\mathrm{RH}^{}$. The final baryon asymmetry in this RHN-dominated case is evaluated by equating $\rho_N^{}=\rho_R^{}$ to get $n_N^{}$ at $T_*^{}$ as
\begin{align}
Y_\mathrm{B}^{}&=\displaystyle \frac{3}{4} c_\mathrm{sph}^{}\epsilon\frac{T_*^{}}{M_1^{}} \label{eq:YBNdom} \\ 
&\simeq 0.26 \sqrt{K} \epsilon\;.\nonumber
\end{align}
The last approximation allows a quick assessment of the generated baryon asymmetry. For observed $Y_\mathrm{B}^{}$ at $\mathcal{O}(10^{-11})$, we need $\sqrt{K} \epsilon$ at $\mathcal{O}(10^{-10})$. It means $\epsilon\simeq 10^{-9}$ for $K\simeq 10^{-2}$, which is much smaller than that typically required by thermal leptogenesis ($\epsilon\simeq 10^{-6}$). 
Following a similar argument, if the RHNs are produced non-relativistically, we have $T_\mathrm{NR}^{}\sim T_\mathrm{RH}^{}$. As long as $T_*^{} < T_\mathrm{NR}^{}$ holds, the same expression in Eq.(\ref{eq:YBNdom}) applies for evaluating the baryon asymmetry. 

When the RHNs decay instantly, i.e., $\Gamma_N^{} \gg \Gamma_\phi^{}$, or equivalently, $T_* \gg T_\mathrm{RH}^{}$, the RHNs decay out before they dominate the Universe, which means instantaneous reheating. The RHN number density is evaluated at $T_\mathrm{RH}^{}$ when they are produced with $n_N^{}=2n_\phi^{}=(\pi^2/15)g_*^{}T_\mathrm{RH}^4/M_\phi^{}$. The final baryon asymmetry in this scenario is
\begin{align}
Y_\mathrm{B}^{}= \displaystyle \frac{3}{2} c_\mathrm{sph}^{}\epsilon \frac{T_\mathrm{RH}^{}}{M_\phi^{}}\;. \label{eq:YBdecay}
\end{align}

The $K\gg1$ limit indicates the Yukawas are strong. In this limit, on the one hand, it is possible to bring the RHNs into thermal equilibrium, although they are produced non-thermally. The implicit condition for this scenario is that $T_*^{} < T_\mathrm{RH}^{}$. This condition is stronger than $M_1^{}<T_\mathrm{RH}^{}$ which is taken as a necessary condition for RHN thermalization~\cite{Giudice:1999fb}.\footnote{Only $M_1^{}<T_\mathrm{RH}^{}$ is thought to be relevant for thermal leptogenesis because it avoids a Boltzmann suppression of the RHN number density in equilibrium (otherwise, the thermalized RHNs are too rare to generate enough lepton asymmetry when they freeze out).} On the other hand, there is also a possibility that the RHNs decay instantly when they are produced, leaving no time for them to form a thermal bath. It happens when $T_*^{} \geq T_\mathrm{RH}^{}$. We can use a simpler but stronger criterion $M_1^{}>T_\mathrm{RH}^{}$ to make a quick assessment. This criterion also implies smaller RHN number density given the same $M_1^{}$ and $M_\phi^{}$, which adds to our understanding that, in addition to having no time to thermalize, the RHN number density is also smaller, providing extra reasoning for not thermalizing in this scenario. In this case, the final baryon asymmetry is evaluated at $T_\mathrm{RH}^{}$ via Eq.~(\ref{eq:YBdecay}).

To sum up, 
we have the following four characteristic limits for RHNs evolution. 
\begin{itemize}
\item The {\it ``RHN dominance"} scenario: characterized by $K \ll 1$ and $T_*^{}<T_\mathrm{NR}^{}$. $T_\mathrm{NR}^{}$ is evaluated as in Eq.~(\ref{eq:Tnr}). For RHNs produced non-relativistically, $T_\mathrm{NR}^{}\simeq T_\mathrm{RH}^{}$. For RHNs produced relativistically, one generally expects $T_*^{}<T_\mathrm{NR}^{}<T_\mathrm{RH}^{}$. So we can conclusively rewrite the second criterion as $T_*^{}<T_\mathrm{RH}^{}$. Depending on whether the RHNs are produced relativistically or not, there will be a radiation-dominated phase before RHN-matter dominated phase or just an RHN-matter-dominated phase. The final baryon asymmetry can be estimated at $T_*^{}$ as in Eq.~(\ref{eq:YBNdom}).
\item The {\it ``instantaneous reheating"} scenario: characterized by $K \ll 1$ and $T_*^{}> T_\mathrm{RH}^{}$, where the latter condition is equivalent to $\Gamma_N^{} \gg \Gamma_\phi^{}$. Although the Yukawa interactions are weak in this scenario, the inflaton-RHN coupling is even weaker. Thus the RHNs decay when they are produced. The final baryon asymmetry can be estimated at $T_\mathrm{RH}$ as in Eq.~(\ref{eq:YBdecay}).
\item The {\it ``thermalized RHNs"} scenario: characterized by $K \gg 1$ and $T_*^{}< T_\mathrm{RH}^{}$. In this scenario, the strong Yukawa interactions bring the RHNs into thermal equilibrium, so it belongs to the thermal leptogenesis. The final baryon asymmetry generally requires solving the Boltzmann equations.
\item The {\it ``strongly non-thermal RHNs"} scenario: characterized by $K \gg 1$ and $T_*^{} > T_\mathrm{RH}^{}$. A necessary condition $M_1^{} > T_\mathrm{RH}^{}$ can be used to make a quick assessment. Although the Yukawa interactions are strong in this scenario, the RHN decay rate is so high that there is no time for the RHNs to thermalize. As mentioned, the RHN number density is also lower, contributing to another obstacle to thermalization. The final baryon asymmetry is evaluated at $T_\mathrm{RH}^{}$ via Eq.~(\ref{eq:YBdecay}).
\end{itemize}
Note that the four limits do not cover all the possibilities. There are cases with an intermediate $K$, for which a Boltzmann investigation is necessary before anything can be said. 

\subsection{Boltzmann equations and numerical results}\label{sec:numeric}

In this subsection, we investigate the evolution processes in the four limits by numerically solving the Boltzmann equations. Throughout this subsection, we will assume that the RHNs are produced as non-relativistic particles. This assumption does not affect leptogenesis -- for the RHNs produced relativistically, decay to Higgs and leptons only happen when they are redshifted to non-relativistic particles. The Boltzmann equations for the inflaton energy density $\rho_\phi^{}$, the RHN energy density $\rho_N^{}$, the radiation energy density $\rho_R^{}$ and the B$-$L number density $n_\mathrm{B-L}^{}$ read~\cite{Giudice:2003jh,Hahn-Woernle:2008tsk}
\begin{align}
\dot{\rho}_\phi^{} &= -3H \rho_\phi^{} - \Gamma_\phi^{} \left(\rho_\phi^{}-\rho_\phi^\mathrm{eq} \right) \;, \nonumber\\
\dot{\rho}_N^{} &= -3H \rho_N^{} + \Gamma_\phi^{} \left(\rho_\phi^{}-\rho_\phi^\mathrm{eq} \right) - \left(\Gamma_N^{}+2\Gamma_{Ss}^{}+4\Gamma_{St}^{} \right)\left(\rho_N^{}-\rho_N^\mathrm{eq} \right) \;, \nonumber\\
\dot{\rho}_R^{} & = -4H \rho_R^{} +  \left(\Gamma_N^{}+2\Gamma_{Ss}^{}+4\Gamma_{St}^{} \right)\left(\rho_N^{}-\rho_N^\mathrm{eq} \right) \;, \nonumber\\
\dot{n}_\mathrm{B-L}^{} &= -3H n_\mathrm{B-L}^{} - \epsilon \Gamma_N^{} \left(n_N^{}-n_N^\mathrm{eq} \right) - \left(W_\mathrm{ID}^{} + \Gamma_{Ss}^{}\frac{\rho_N^{}}{\rho_N^\mathrm{eq}}+2\Gamma_{St}^{}\right) n_\mathrm{B-L}^{} \;, 
\end{align}

where $\Gamma_{Ss}^{}, \Gamma_{St}^{}$ represent $\Delta \mathrm{L}=1$ scattering rates involving the top quark and gauge bonsons (summed in $s-$ and $t-$ channels respectively), $W_\mathrm{ID}$ denotes the washout contribution from the RHN inverse decay,  $W_\mathrm{ID}^{}=\Gamma_\mathrm{ID}^{}/2=\Gamma_N^{} n_N^\mathrm{eq}/n_l^\mathrm{eq}/2$ with $n_N^\mathrm{eq}$ and $n_l^\mathrm{eq}$ being the RHN equilibrium number density and the lepton doublet equilibrium number density. We include the inverse decay term for the inflaton, which can be neglected if the reheating temperature is smaller than the inflaton mass, so the inflaton cannot enter thermal equilibrium.

We neglect $\Delta \mathrm{L}=2$ scattering processes as they are subdominant for the considered scale. There are inflaton-RHN scattering processes induced by the inflaton-RHN coupling. A detailed study on such scalar-RHN scatterings and their effect on leptogenesis is given in Ref.~\cite{AristizabalSierra:2014uzi}. We check that these scattering rates are quite suppressed if the scalar is heavier than the RHNs, which is our considered cases.

In practice, it is convenient to scale out the expansion effect. We follow the literature~\cite{Chung:1998rq,Hahn-Woernle:2008tsk} and introduce
\begin{align}
E_\phi^{} = \rho_\phi^{} a^3,~E_N^{} = \rho_N^{} a^3,~N_\mathrm{B-L}^{} = n_\mathrm{B-L}^{} a^3,~E_R^{} = \rho_R^{} a^4\;,\label{eq:covDef}
\end{align}
where $a$ is the scale factor. We further define $y=a/a_\mathrm{I}^{}$ and choose the initial value of the scale factor as $a_\mathrm{I}^{}=1$. With these quantities, the Boltzmann equations can be rewritten into
\begin{align}
\displaystyle \frac{\mathrm{d} E_\phi^{}}{\mathrm{d}y} &= - \frac{\Gamma_\phi^{}}{Hy} \left(E_\phi^{} - E_\phi^\mathrm{eq} \right) \;, \nonumber\\
\displaystyle \frac{\mathrm{d} E_N^{}}{\mathrm{d}y} &= \frac{\Gamma_\phi^{}}{Hy} \left(E_\phi^{} - E_\phi^\mathrm{eq} \right) -\frac{1}{Hy}\left(\Gamma_N^{}+2\Gamma_{Ss}^{}+4\Gamma_{St}^{} \right) \left(E_N^{} - E_N^\mathrm{eq} \right) \;, \nonumber\\
\displaystyle \frac{\mathrm{d} E_R^{}}{\mathrm{d}y} &= \frac{1}{Hy}\left(\Gamma_N^{}+2\Gamma_{Ss}^{}+4\Gamma_{St}^{} \right) \left(E_N^{} - E_N^\mathrm{eq} \right) \;, \nonumber\\
\displaystyle \frac{\mathrm{d} N_\mathrm{B-L}^{}}{\mathrm{d}y} &= -\frac{\epsilon\Gamma_N^{}}{Hy} \left(N-N^\mathrm{eq}_{} \right) - \frac{1}{Hy} \left(W_\mathrm{ID}^{} + \Gamma_{Ss}^{}\frac{\rho_N^{}}{\rho_N^\mathrm{eq}}+2\Gamma_{St}^{}\right)  N_\mathrm{B-L}^{}\;. \label{eq:BE}
\end{align}

The Hubble expansion rate is determined from the Friedmann equation assuming a zero curvature,
\begin{align}
H^2= \displaystyle \frac{8\pi}{3 M_\mathrm{pl}^2} \left( \rho_\phi^{} + \rho_N^{} + \rho_R^{} \right) = \frac{8\pi}{3 M_\mathrm{pl}^2 y^4} \left( E_\phi^{}y+ E_N^{}y+E_R^{} \right) \;.
\end{align}
Temperature is defined once there is a thermal ensemble of radiation and is expressed as
\begin{align}
T=\displaystyle \left( \frac{30}{\pi^2 g_*^{}} E_R^{}\right)^{1/4}  \frac{1}{y} \;, \label{eq:Tvsy}
\end{align}
where we use $\rho_R^{} =(\pi^2/30)g_*^{} T^4  = E_R^{}/a^4$ and $a_\mathrm{I}^{}=1$.

For our numerical studies, we use $\{ K, M_1^{}, M_\phi^{}, T_\mathrm{RH}^{}\}$ as inputs and adjust their values such that they meet the conditions defining each of the four scenarios. The decay parameter $K$ characterizes the strength of the Yukawa interactions and also the washout effects given $M_1^{}$ and can be viewed as a measure of efficiency. $T_\mathrm{RH}^{}$ is the physical reheating temperature when the RHNs decay instantly, and it parameterizes the strength of the inflaton-RHN coupling. 

The final baryon asymmetry can be expressed in terms of $n_\mathrm{B-L}^{}/n_\gamma^\mathrm{eq}$ as~\cite{Buchmuller:2004nz, Davidson:2008bu}
\begin{align}
Y_\mathrm{B}^{} \simeq \frac{c_\mathrm{sph}^{}}{7.04f} \frac{n_\mathrm{B-L}^{}}{n_\gamma^\mathrm{eq}} \;,
\end{align}
where $f = 2387/86$ is the dilution factor from the onset of leptogenesis till the recombination, $c_\mathrm{sph}^{}$ accounts for the B$-$L to B conversion, and $7.04 \simeq s/n_\gamma^{}$ evaluated at present. The last factor $n_\mathrm{B-L}^{}/n_\gamma^\mathrm{eq}$ relates to the $N_\mathrm{B-L}^{}$ (and $E_R^{}$) calculated in the Boltzmann equations as
\begin{align}
\frac{n_\mathrm{B-L}^{}}{n_\gamma^\mathrm{eq}}= \displaystyle \frac{\pi^{7/2} g_*^{3/4}}{2 \times 30^{3/4} \zeta(3) E_R^{3/4}} N_\mathrm{B-L}^{} \;,
\label{eq:NBmL}
\end{align}
where $n_\gamma^\mathrm{eq}=(2\zeta(3) T^3)/\pi^2$, and we use $T$ in Eq.~(\ref{eq:Tvsy}).
To generate an observed value of the baryon asymmetry, i.e., $Y_\mathrm{B}^{}=\left(8.72\pm 0.08\right) \times 10^{-11}$~\cite{Planck:2018vyg}, $n_\mathrm{B-L}^{}/n_\gamma^\mathrm{eq}$ should be around $\mathcal{O}(10^{-8})$.

\subsubsection{$K\ll 1$ scenarios}

\begin{figure}[t!]
\centering
\includegraphics[width=.9\textwidth]{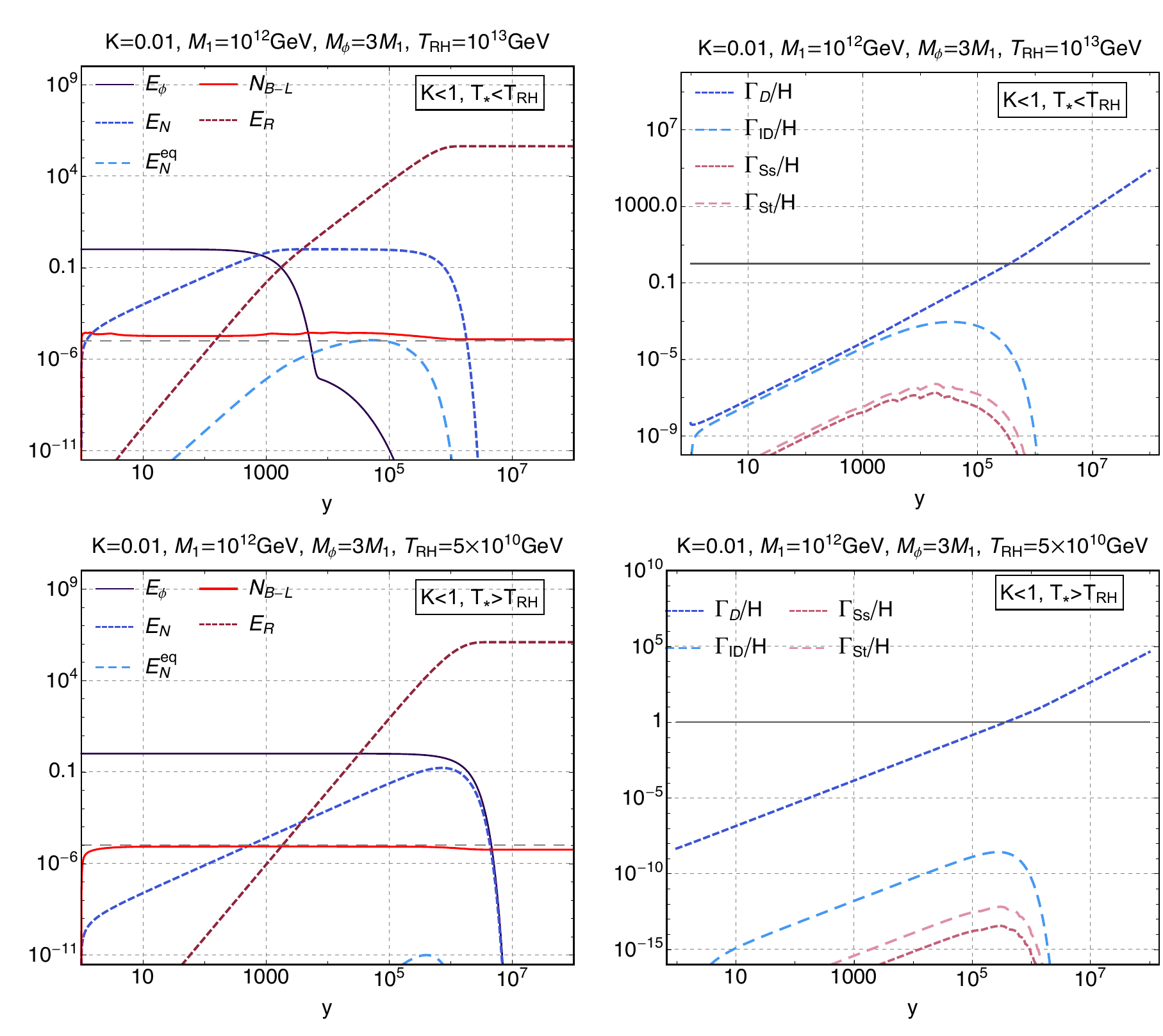}
\caption{The evolution of the rescaled energy densities (left column)  and the interaction rates (right column) as a function of the rescaled dimensionless scale factor $y=a/a_\mathrm{I}^{}$ in the cases with $K=0.01$. The energy densities are normalized to the initial inflation energy $E_\phi^0$. We fix $|\epsilon|=10^{-6}$. The horizontal gray dashing lines are the analytic estimates. }
\label{fig:weakWO}
\end{figure}

We show the numerical results in the two $K\ll 1$ scenarios in Fig.~\ref{fig:weakWO}. The four input parameters are arranged to meet the conditions
$T_*^{} < T_\mathrm{RH}^{}$ or $T_*^{} > T_\mathrm{RH}^{}$. We find the evolution processes as expected: In both scenarios, the RHNs do not enter the thermal equilibrium; The $N_\mathrm{B-L}$ lines are rather flat, indicating the washout effects are small for $K\ll 1$.

The two rows of Fig.~\ref{fig:weakWO} are calculated with the same initial values, and the only difference in input is the values of $T_\mathrm{RH}^{}$. We start with the same amount of $E_\phi^{}$ and vanishingly small other quantities. As $T_\mathrm{RH}^{}$ characterizes $\Gamma_\phi^{}$, a larger $T_\mathrm{RH}^{}$ leads to a sooner inflaton decay, resulting in an earlier generation of RHNs and hence radiation. $E_N^\mathrm{eq}$ is mainly a reflection of thermal bath temperature given the same $M_N^{}$. The noticeable differences in $E_N^\mathrm{eq}$ result from the difference in $E_R^{}$, from which we define the thermal bath temperature in Eq.~(\ref{eq:Tvsy}). We also observe the higher $T_\mathrm{RH}^{}$ scenario experiences a slightly stronger washout as a result of the slightly higher thermal bath temperature.

We draw the analytical estimates in these benchmark scenarios with gray dashing lines in the left column plots of Fig.~\ref{fig:weakWO} for comparison. They roughly agree with the numeric results. The analytic estimates from Eq.~(\ref{eq:YBNdom}) and Eq.~(\ref{eq:YBdecay}) give $Y_\mathrm{B}^{} \simeq 2.6\times 10^{-8}$ and $Y_\mathrm{B}^{} \simeq 8.9\times 10^{-9}$ in the two scenarios, while the numeric results render $Y_\mathrm{B}^{} \simeq 2.3\times 10^{-8}$ and $Y_\mathrm{B}^{} \simeq 1.0\times 10^{-8}$. The slightly larger $Y_\mathrm{B}^{}$ in the {\it instantaneous reheating} scenario is caused by the small initial values we set for the Boltzmann system and is not physical.

\subsubsection{$K\gg 1$ scenarios}\label{sec:strongWO}

\begin{figure}[t!]
\centering
\includegraphics[width=.9\textwidth]{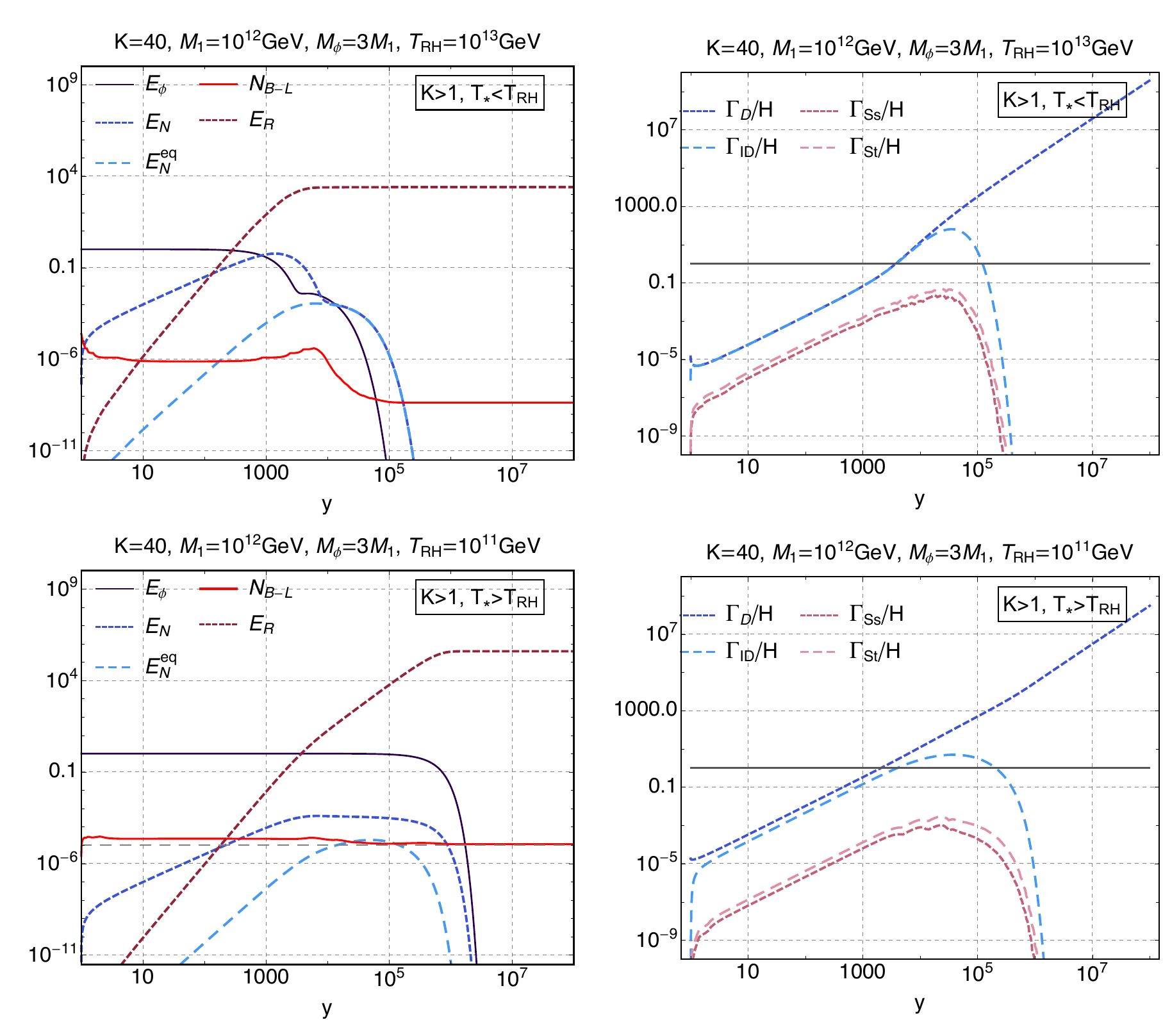}
\caption{The evolution of the rescaled energy densities (left column)  and the interaction rates (right column) as a function of the rescaled dimensionless scale factor $y=a/a_\mathrm{I}^{}$ in the cases with $K=40$. The energy densities are normalized to the initial inflation energy $E_\phi^0$. We fix $|\epsilon|=10^{-6}$. The horizontal gray dashing line is the analytic estimate.}
\label{fig:strongWO}
\end{figure}

We also show the numerical results in the two $K\gg 1$ scenarios in Fig.~\ref{fig:strongWO}. The value of $K=40$ is chosen as follows. From Eq.~(\ref{eq:mstar}), one has $m_*^{} \simeq 1.1 \times 10^{-3}$ eV. Setting $\tilde{m}_1^{}$ to be in the range given by the two squared-mass differences, i.e., $[0.0086,0.05]$ eV, one finds $K \in [7.8,45]$. 

The general evolution processes fit our expectations. In $T_*^{}<T_\mathrm{RH}^{}$ case, we find $E_N^{}$ follows closely to $E_N^\mathrm{eq}$, indicating that the RHNs enter thermal equilibrium. In $T_*^{}>T_\mathrm{RH}^{}$ case, the RHNs decay before they can enter the thermal equilibrium. They decay out almost at the same time the inflaton decays out. 

Regarding inputting variables, the two benchmark scenarios differ in $T_\mathrm{RH}^{}$. Again, we see that a higher $T_\mathrm{RH}^{}$ leads to sooner inflaton decay, an earlier generation of the RHNs and the radiation, and a higher thermal bath temperature that leads to a stronger washout. The analytic estimation from Eq.~(\ref{eq:YBdecay}) gives $Y_\mathrm{B}^{} \simeq 1.8\times 10^{-8}$, which is shown with a gray dashing line in comparison with the numeric result $Y_\mathrm{B}^{} \simeq 2.1\times 10^{-8}$.

\subsection{Final efficiency}\label{sec:efficiency}

We introduce the final efficiency factor $\kappa_f^{}$ to parameterize the contribution to the baryon asymmetry that is independent on the CP asymmetry $\epsilon$, in the same way as was done for thermal leptogenesis~\cite{Buchmuller:2004nz} to allow for an easy comparison
\begin{align}
\frac{n_\mathrm{B-L}^{}}{n_\gamma^\mathrm{eq}}=-\frac{3}{4} \epsilon \kappa_f ^{}\;,
\end{align}
where the factor $3/4$ comes from $n_N^\mathrm{eq}/n_\gamma^\mathrm{eq}$ and the deviation to the thermal number density is parameterized in $\kappa_f^{}$. The final efficiency factor $\kappa_f^{}$ is independent of the CP asymmetry but depends on all the four input parameters $\{ K, M_1^{}, M_\phi^{}, T_\mathrm{RH}^{}\}$. As we consider only non-relativistically produced RHNs, we fix $M_\phi^{}=3M_1^{}$ for definiteness.
Eq.~(\ref{eq:Trh}) allows us to re-express the inflaton-RHN coupling $y_N^{}$ in terms of $T_\mathrm{RH}^{}$ and $M_\phi^{}$. As a result, the perturbativity bound on the inflaton-RHN coupling $y_N^{} < 4\pi$ can be translated into an upper bound of $T_\mathrm{RH}^{}$ for a given $M_\phi^{}$
\begin{align}
T_\mathrm{RH}^{} < \sqrt{4\pi} \left(\frac{45}{4\pi^3 g_*^{}} \right)^{1/4} \sqrt{M_\phi^{} m_\mathrm{pl}^{}}\;.\label{eq:Trh-bound}
\end{align}
Assuming $M_\phi^{}=3M_1^{}$, Eq.~(\ref{eq:Trh-bound}) can be re-written as a lower bound on $M_1^{}$ given $T_\mathrm{RH}^{}$ 
\begin{align}
M_1 > \frac{1}{12\pi} \left(\frac{4\pi^3 g_*^{}}{45} \right)^{1/2} \frac{T_\mathrm{RH}^2}{m_\mathrm{pl}^{}}  \;.
\end{align}
For examples, the bounds are $T_\mathrm{RH}^{}<5.2\times 10^{15}$ GeV for $M_1^{}=10^{12}$ GeV; $M_1^{}>3.7\times 10^4$ GeV for $T_\mathrm{RH}^{}=10^{12}$ GeV. 

\begin{figure}[t!]
\centering
\includegraphics[width=.5\textwidth]{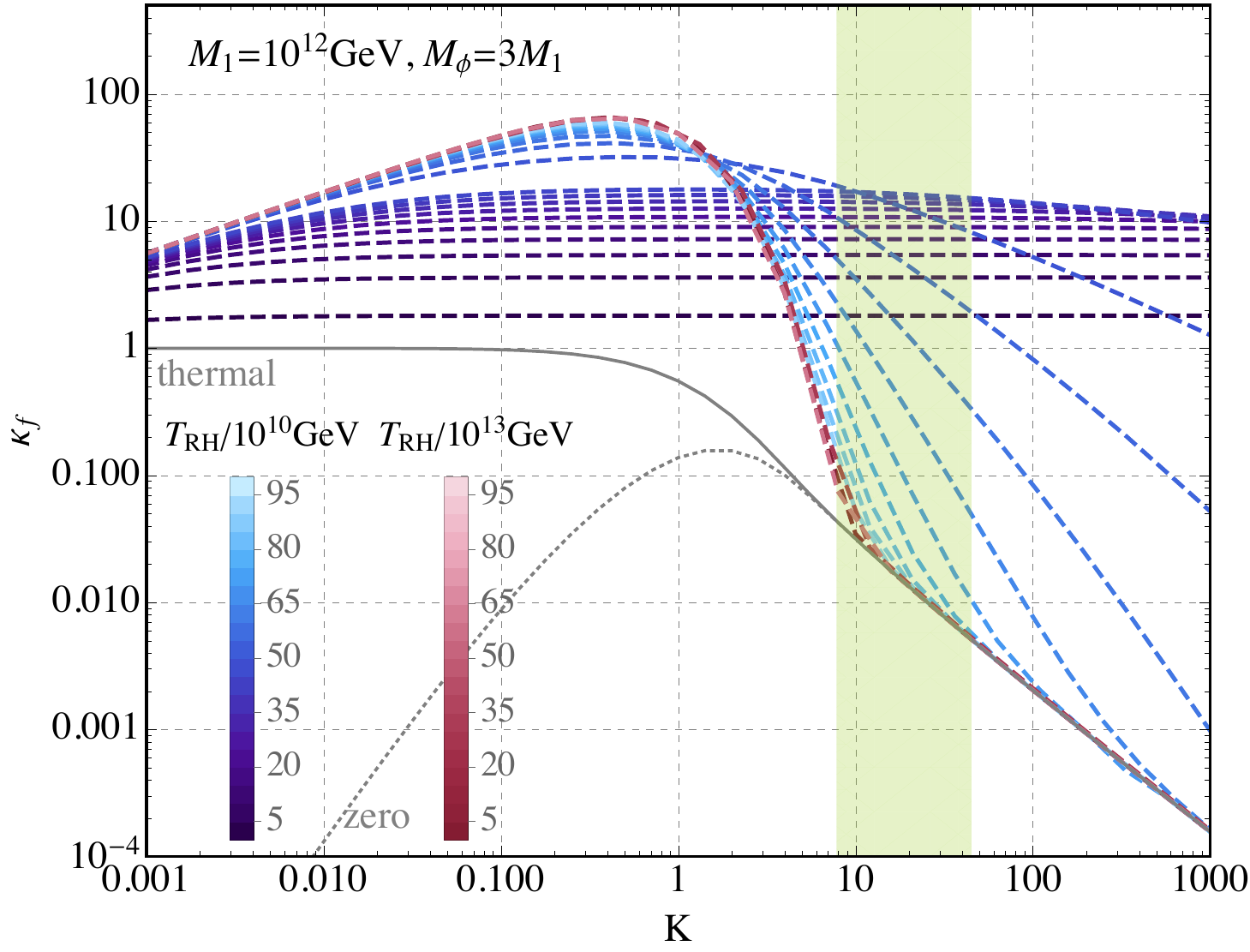}
\caption{The final efficiency factor as a function of the decay parameter $K$ with different $T_\mathrm{RH}^{}$. Also shown are the thermal initial abundance case (gray solid line) and zero initial abundance case (gray dotted line) for comparison. The two squared-mass differences observed in neutrino oscillation experiments give the two vertical dashed lines. We fix $M_\phi^{}=3M_1^{}$ and $M_1^{}=10^{12}$ GeV.}
\label{fig:kk_compare}
\end{figure}

Including the $\kappa_f ^{} - K$ plot to compare with thermal leptogenesis directly is informative. We numerically investigate the effects of varying $T_\mathrm{RH}^{}$ and show the results in Fig.~\ref{fig:kk_compare}.\footnote{In calculation here, we neglect the scattering processes as their contributions are subdominant compared with the decay and inverse decay.}A direct observation tells that non-thermal leptogenesis generally has a larger efficiency factor. The four limits lie on the left and right sides, say, $K \leq 0.01$ and $K\geq 10$. On the left, the blue lines from the top correspond to the RHN dominance scenario with $T_*^{}<T_\mathrm{RH}^{}$. The few dark blue lines right above the thermal line are the instantaneous reheating scenario with $T_*^{}<T_\mathrm{RH}^{}$. On the right, the dark red lines feature $T_*^{}<T_\mathrm{RH}^{}$ and become identical to the thermal line for large enough $K$, indicating the RHNs in this scenario can thermalize. From blue to dark blue lines, we see the final efficiency first grows, then decreases with a growing $M_1^{}/T_\mathrm{RH}^{}$. The growing period actually corresponds to the intermediate cases. If the Yukawa interactions are stronger, they can be brought into equilibrium. The decreasing period corresponds to the true {\it strongly non-thermal RHNs} scenario, where the RHNs cannot be thermalized even if $K$ gets stronger. Featuring a relatively low reheating temperature, the inflaton decay is delayed in this strongly non-thermal RHNs scenario, leading to an RHN density (hence the radiation density) not large enough to be thermalized. As such, the lower the reheating temperature, the smaller the final efficiency factor. The intermediate range of $K$, i.e., $0.01<K<10$, generally has a larger efficiency factor than the thermal case to a maximum of $\kappa_f^{}|_\mathrm{max}^{} \simeq 70$.

The {\it instantaneous reheating} scenario ($K\ll 1, T_*^{}>T_\mathrm{RH}^{}$) and the {\it strongly non-thermal RHNs} scenario are represented by the same rather flat lines in the $\kappa_f ^{} - K$ plane. The larger $T_*^{}/T_\mathrm{RH}^{}$, the flatter the lines are. The final baryon asymmetry in both scenarios can be evaluated through Eq.~(\ref{eq:YBdecay}), where $K$ dependence is implicit in $\epsilon$. As a result, we conclude that as long as $T_*^{}\gg T_\mathrm{RH}^{}$ hold, one can use Eq.~(\ref{eq:YBdecay}) to estimate the baryon asymmetry, regardless of the $K$ value. On the other hand, when $T_*^{} \leq T_\mathrm{RH}^{}$, only a small $K$ limit gives a non-thermal scenario, i.e., the {\it RHN dominance} limit, that analytic estimate in Eq.~(\ref{eq:YBNdom}) works. For other values of $K$, a Boltzmann investigation is necessary to evaluate the final baryon asymmetry.

\subsection{Neutrino mass bound and the reheating temperature}\label{sec:numass}

There exists an upper bound on the CP asymmetry generated by the RHNs decay, i.e., the Davidson-Ibarra bound~\cite{Davidson:2002qv}, 
\begin{align}
|\epsilon_1^{}| \leq \frac{3}{16\pi} \frac{M_1^{} m_\mathrm{atm}}{v^2} \;,
\end{align}
where $m_\mathrm{atm}$ is the root of the squared-mass difference measured first in atmospheric oscillation experiments. This bound can be transformed to a lower bound on the RHN mass~\cite{Buchmuller:2004nz,Hahn-Woernle:2008tsk}
\begin{align}
M_1^{} &\geq 6\times 10^8 ~\mathrm{GeV} \left(\frac{Y_\mathrm{B}^{}}{8\times 10^{-11}} \right) \left( \frac{0.05~\mathrm{eV}}{m_\mathrm{atm}^{}}\right) \kappa_f^{-1} (K) \nonumber\\
&\geq 6.28 \times 10^8 ~\mathrm{GeV} \left( \frac{0.05~\mathrm{eV}}{m_\mathrm{atm}^{}}\right) \kappa_f^{-1} (K) \;,
\end{align}
where in the last inequality we use the $3\sigma$ range of the observed $Y_\mathrm{B}^{}$ in Eq.~(\ref{eq:BAU_obs})~\cite{Planck:2018vyg}.

With the $\kappa_f^{}$ - $K$ relation determined in the previous subsection, we can find the lower bound on the RHN mass $M_1^{}$. We show the results in Fig.~\ref{fig:mkflv}, where the thermal initial abundance line (orange dashed) and zero initial abundance line (gray dashed) are adopted from Ref.~\cite{Giudice:2003jh}, and the red solid line is given by including flavor effect (to be discussed in the next subsection). All the region above the black (red) solid line is viable for the unflavored (flavored) non-thermal leptogenesis.  The right-handed side of the blue dashed line is allowed only when $M_1^{} > T_\mathrm{RH}^{}$, and the left-handed side can happen for both. Compared with the cases that the RHNs can thermalize, the {\it strongly non-thermal} scenario opens up parameter space at the large $K$ region greatly, which is also the oscillation-preferred region. We find $M_1^{} \geq 10^7$ GeV for non-thermal leptogenesis in general and $M_1^{} \geq 2\times 10^7$ GeV in the oscillation-preferred region. The results are compatible with literature~\cite{Giudice:2003jh,Hahn-Woernle:2008tsk}.

The reheating temperature in thermal leptogenesis gets a similar lower bound to that of $M_1^{}$ since we need it to be higher or equal to $M_1^{}$ such that the RHNs can be in thermal equilibrium with a not-suppressed abundance. In thermal leptogenesis, the RHNs are not required and even encouraged to deviate from equilibrium so $T_\mathrm{RH}^{}$ is not constrained given the lower limit of $M_1^{}$. We have seen in Fig.~\ref{fig:kk_compare} that when $M_1^{} = 100 T_\mathrm{RH}^{}$ in the whole $K$ range, non-thermal leptogenesis has a final efficiency larger than that of thermal one. Noticeably, when $M_1^{} \gg T_\mathrm{RH}^{}$, the final efficiency is insensitive to $K$ so we can estimate the final baryon asymmetry with Eq.~(\ref{eq:YBdecay}). For a CP asymmetry $\epsilon \sim \mathcal{O}(10^{-6})$, one can have enough baryon asymmetry for $M_1^{}/T_\mathrm{RH}^{} \sim 10^4$. Put the lower limit on $M_1^{}$ in, we see that $T_\mathrm{RH}^{}$ can be lowered to $10^3$ GeV.

\subsection{Flavor effects}\label{sec:flavor}

\begin{figure}[t!]
\centering
\includegraphics[width=.5\textwidth]{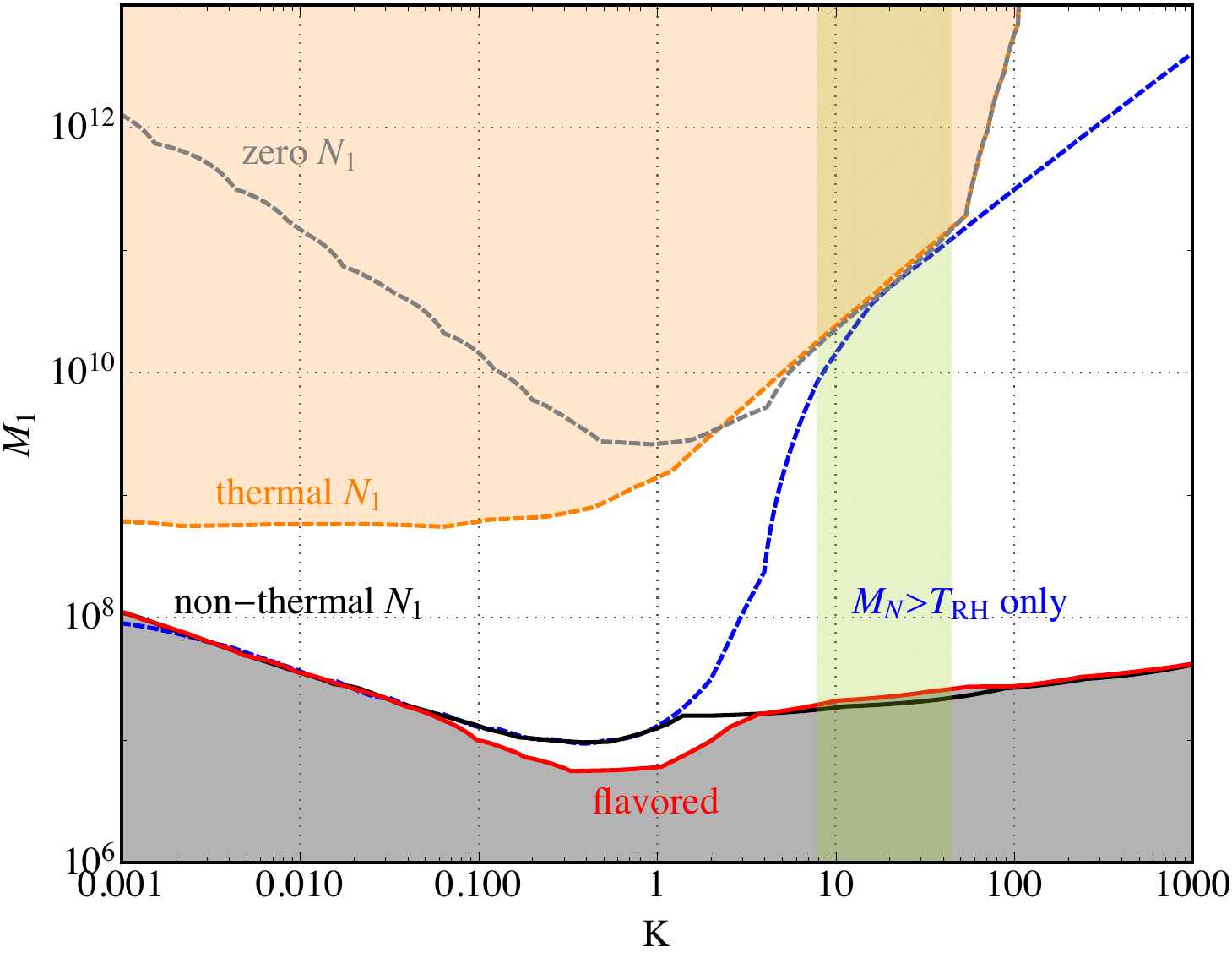}
\caption{The lower limit on $M_1^{}$ as a function of the decay parameter $K$ for the unflavored treatment (the black solid line) and the two-flavored one (the red solid line) in non-thermal leptogenesis. The dashed blue line markers the boundary of whether the non-thermally produced RHNs can be thermalized. On the right of the dashed blue line, the RHNs cannot be thermalized, and this region only works for $M_1^{} > T_\mathrm{RH}^{}$. Also shown are the thermal initial abundance case (orange dashed line) and zero initial abundance case (gray dashed line) adopted from Ref.~\cite{Giudice:2003jh} for comparison. The two squared-mass differences observed in neutrino oscillation experiments give light green bands.}
\label{fig:mkflv}
\end{figure}

When the charged lepton Yukawa interactions enter equilibrium, flavor basis matters, and the unflavored treatment may be inappropriate. To be definite, let us consider only the $\tau$ Yukawa interaction entering the equilibrium to illustrate the flavor effect. The interaction rate is $\Gamma_\tau^{} \simeq 5\times 10^{-3} h_\tau^2T$ with $h_\tau^{}$ being the $\tau$ Yukawa coupling constant. Similar to thermal leptogenesis, when $\Gamma_\tau^{}>\Gamma_\mathrm{ID}^\tau>H$, the $\tau$ flavor is distinguishable and one should separate the $N_\mathrm{B-L}^{}$ equation into two with different flavor contributions. When non-thermal leptogenesis is considered, there is also the possibility that $\Gamma_\tau^{}>H>\Gamma_\mathrm{ID}^\tau$, we will check the effect in this regime. We work in the two-flavored regime and write the Boltzmann equations as
\begin{align}
\displaystyle \frac{\mathrm{d} E_\phi^{}}{\mathrm{d}y} &= - \frac{\Gamma_\phi^{}}{Hy} \left(E_\phi^{} - E_\phi^\mathrm{eq} \right) \;, \nonumber\\
\displaystyle \frac{\mathrm{d} E_N^{}}{\mathrm{d}y} &= \frac{\Gamma_\phi^{}}{Hy} \left(E_\phi^{} - E_\phi^\mathrm{eq} \right) -\frac{\Gamma_N^{}}{Hy} \left(E_N^{} - E_N^\mathrm{eq} \right) \;, \nonumber\\
\displaystyle \frac{\mathrm{d} E_R^{}}{\mathrm{d}y} &= \frac{\Gamma_N^{}}{H} \left(E_N^{} - E_N^\mathrm{eq} \right) \;, \nonumber\\
\displaystyle \frac{\mathrm{d} N_{\Delta_\tau}^{}}{\mathrm{d}y} &= -\frac{\epsilon_\tau^{}\Gamma_N^{}}{Hy} \left(N-N^\mathrm{eq}_{} \right) - \frac{P_\tau^{} W_\mathrm{ID}^{}}{Hy} \sum_{\alpha=\tau,\tau^\perp_{}} C_{\tau \alpha}^{} N_{\Delta_\alpha}^{}\;, \nonumber \\
\displaystyle \frac{\mathrm{d} N_{\Delta_{\tau^\perp}}^{}}{\mathrm{d}y} &= -\frac{\epsilon_{\tau^\perp}^{}\Gamma_N^{}}{Hy} \left(N-N^\mathrm{eq}_{} \right) - \frac{P_{\tau^\perp}^{} W_\mathrm{ID}^{}}{Hy} \sum_{\alpha=\tau,\tau^\perp_{}} C_{{\tau^\perp} \alpha}^{} N_{\Delta_\alpha}^{}\;,\label{eq:BEflavored}
\end{align}
where the charge asymmetry in single flavor is $\Delta_\alpha^{} =\mathrm{B}/3 -\mathrm{L}_\alpha^{}$ and $\tau^\perp_{}$ is a combination of the $e$ and $\mu$ flavor with $P_\alpha^{} \equiv |\langle l_1^{} | l_\alpha^{} \rangle |^2$ projecting the charged lepton generated by RHN decay ($l_1^{}$) into flavor basis ($l_\alpha^{}$). The total asymmetry is $N_\mathrm{B-L}^{}=N_{\Delta_\tau}^{}+N_{\Delta_{\tau^\perp}}^{}$. The CP asymmetry in flavor $\alpha$ is
\begin{align}
\epsilon_\alpha^{} = \frac{1}{2} \left( P_\alpha^{} + \overline{P}_\alpha^{} \right) \epsilon + \frac{1}{2} \Delta P_\alpha^{}\;,
\end{align}
where $\overline{P}_\alpha^{}$ is the antilepton corresndance of $P_\alpha^{}$ and $\Delta P_\alpha^{} \equiv P_\alpha^{} - \overline{P}_\alpha^{}$. $\Delta P_\alpha^{}=0$ at the tree level and is nonzero when loop effects are included. This term is crucial to get deviations from the unflavored treatment~\cite{Nardi:2006fx,Dev:2017trv}. In our numeric treatment, we use $P_\alpha^{}/\epsilon \propto \sqrt{K_\alpha^{}/K}$, where $K_\alpha^{}=P_\alpha^{} K$~\cite{Davidson:2008bu}. The $C$ matrix relates the asymmetry in each charged lepton flavor $N_{\Delta_{l_\alpha}}^{}$ with the charge asymmetry in the flavor $N_{\Delta_\alpha}^{}$, $N_{\Delta_{l_\alpha}}^{}=\sum_{\beta} C_{\alpha \beta} N_{\Delta_\beta}^{}$. Here we use the one including both the sphaleron-induced lepton flavor mixing and the Higgs asymmetry, which is given by~\cite{Dev:2017trv}
\begin{align}
C= \left(
\begin{array}{cc}
 \displaystyle \frac{581}{589} & \displaystyle \frac{104}{589}\vspace{0.3cm} \\
 \displaystyle \frac{194}{589} & \displaystyle \frac{614}{589}\\
\end{array} \right)\;.
\end{align}

\begin{figure}[t!]
\centering
\includegraphics[width=.9\textwidth]{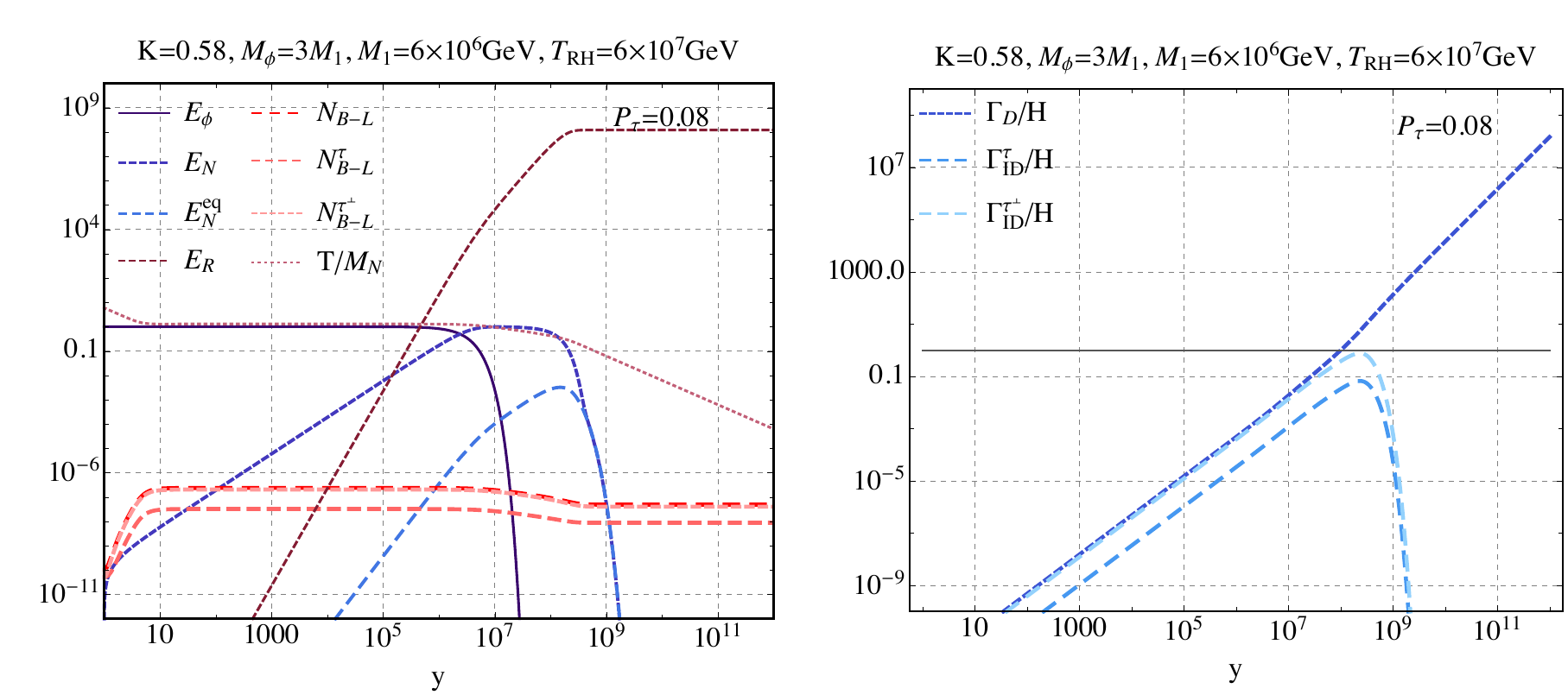}
\caption{The evolution of the rescaled energy densities (left column)  and the decay (inverse decay) rates (right column) as a function of the rescaled dimensionless scale factor $y=a/a_\mathrm{I}^{}$ when flavor effects are included. The energy densities are normalized to the initial inflation energy $E_\phi^0$.}
\label{fig:evlflv}
\end{figure}

We perform a parameter space scan and present the results in Fig.~\ref{fig:mkflv}. The lower bound on $M_1^{}$ is lowered maximally to $6\times 10^6$ GeV when $0.1\leq K \leq3$, whereas in the other region, identical to the unflavored treatment. For $K< 0.1$, the inverse decay rates in both flavors are much lower than the Hubble expansion rate. It corresponds to $\Gamma_\tau^{}>H>\Gamma_\mathrm{ID}^\tau$. Since the inverse decay rates in both flavors are negligible, working in the flavor basis will not make much difference compared with the unflavored basis. For $K>3$, there exist two possibilities. If $M_1^{} \leq T_\mathrm{RH}^{}$, the RHNs will be thermalized eventually and the $M_1^{}$-$K$ line resume the thermal one. If $M_1^{} \gg T_\mathrm{RH}^{}$, the RHNs are not abundantly produced due to the delayed inflaton decay. The number of RHNs and hence the radiation is insufficient to bring the RHNs into equilibrium, and the $M_1^{}$-$K$ line resumes the non-thermal line in the unflavored treatment. We also find that the lower limit of $M_1^{}$ is insensitive to the value of $P_\alpha^{}$. In other words, for any value of $P_\alpha^{}$, the lower limit of $M_1^{}$ can be achieved if we allow other parameters to vary.

We plot the evolution of the various densities in Fig.~\ref{fig:evlflv} for one of the points approaching the lower limit of $M_1^{}$. We see that the total B$-$L density $N_\mathrm{B-L}^{}$ follows $N_\mathrm{B-L}^{\tau^\perp}$ while the other flavor contribution $N_\mathrm{B-L}^{\tau}$ is negligible. The RHNs dominate the energy for a while, indicating that we are in the {\it RHN dominance} scenario. Looking from the rate plot on the right, we see a similar situation to the case that maximizes $\kappa_f^{}$, i.e., $y_\mathrm{in}^{}=y_\mathrm{out}^{}=y_\mathrm{max}^{}$ for $\Gamma_\mathrm{ID}^{\tau^\perp}=2 W_\mathrm{ID}^{\tau^\perp}=2 P_{\tau^\perp}^{} W_\mathrm{ID}^{}$, where $y_\mathrm{in}^{}, y_\mathrm{out}^{}$ mark the time (scale) $\Gamma_\mathrm{ID}^{\tau^\perp}$ enters and leaves equilibrium, $y_\mathrm{max}^{}$ denotes the time (scale) $\Gamma_\mathrm{ID}^{\tau^\perp}$ reach its maximal value.

\subsection{Summary}\label{sec:sum}

\begin{figure}[t!]
\centering
\includegraphics[width=\textwidth]{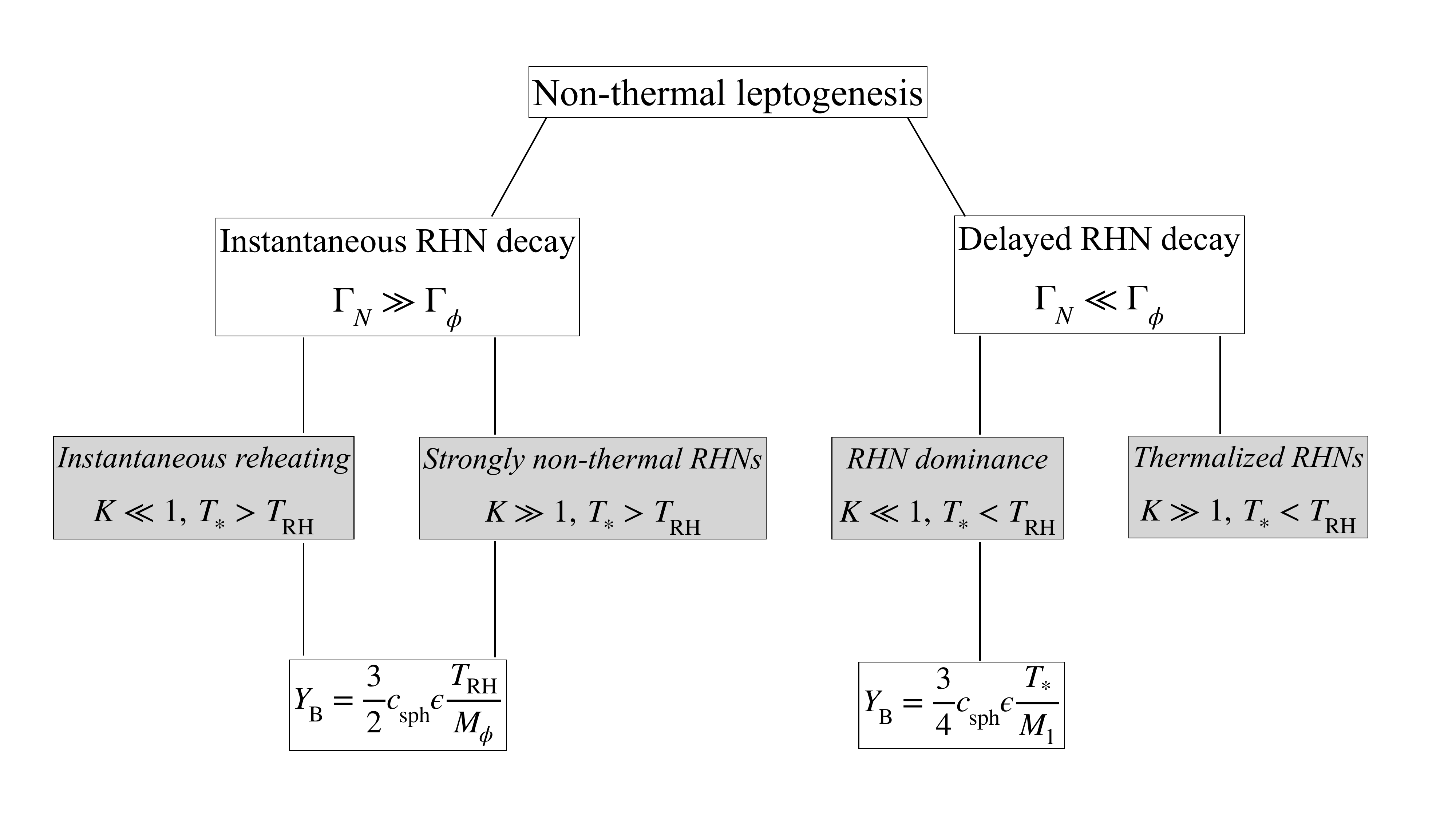}
\caption{The classification of non-thermal leptogenesis. The four characteristic limits are shown in grey area with working conditions to identify them. They belong to two general categories. Analytic estimates are given for the three of them that the RHNs are not thermalized.}
\label{fig:ntClass}
\end{figure}

Returning to the two questions we proposed at the beginning of Sec.~\ref{sec:nt}, we summarize the answers in the tree diagram in Fig.~\ref{fig:ntClass}. There are in general two categories. The one with $\Gamma_N^{} \gg \Gamma_\phi^{}$ features an instantaneous RHN decay, which instantly reheats the Universe. The other one with $\Gamma_N^{} \ll\Gamma_\phi^{}$ features a delayed RHN decay, and an RHN dominance follows. We prefer not to use the name ``instantaneous reheating" for $\Gamma_N^{} \gg \Gamma_\phi^{}$ as we save it for the subcategory. We put the four characteristic limits under these two categories and also list the working conditions and the analytic estimations that apply. For intermediate cases, a Boltzmann investigation is necessary.

Several comments are in order.
\begin{itemize}
\item The characteristic evolution process is the appearance of an RHN matter-dominated era. It happens when $T_*^{}<T_\mathrm{RH}^{}$ for both a large and a small $K$. In the case of a large $K$, the RHN matter dominance happens before they go to thermal equilibrium. For a small $K$, the RHN matter dominance happens as long as the RHNs become non-relativistic since their energy redshifts slower than the radiation. 
\item Regarding the characteristic quantities, $\Gamma_N^{}$ versus $\Gamma_\phi^{}$ working in combination with different limits of $K$ allow us to identify the four scenarios. In practice, it is convenient to use the working conditions expressed in terms of $T_*^{}$ and $T_\mathrm{RH}^{}$ for each scenario to make a quick estimation\footnote{According to the definition of $T_*^{}$ in Eq.~(\ref{eq:Tstar}), converting $\Gamma_N^{}$ versus $\Gamma_\phi^{}$ relation to $T_*^{}$ versus $T_\mathrm{RH}^{}$ hold for negligible dilation factor. This approximation works well for $M/T \geq 1$, while for $M_1^{}/T \sim \mathcal{O}(0.01)$, the approximation causes an error at $\mathcal{O}(0.1)$~\cite{Buchmuller:2004nz}.}.
\item Three of the four limits (except the {\it thermalized RHNs} limit) have larger final efficiency factors than that of thermal leptogenesis. Analytical estimates for the final baryon asymmetry for all three non-thermal limits exist. As discussed in Sec.~\ref{sec:efficiency}, Eq.~(\ref{eq:YBdecay}) hold for all values of $K$ in the $T_*^{}\gg T_\mathrm{RH}^{}$ limit. While when $T_*^{}\leq T_\mathrm{RH}^{}$, only a small $K$ gives the analytic estimate in Eq.~(\ref{eq:YBNdom}). For the rest scenarios, Boltzmann equations are needed. 
\item The {\it strongly non-thermal RHNs} scenario is very interesting as it seems counter-intuitive at first sight. ``Strong" refers to large $K$, which implies strong Yukawa for given $M_1^{}$. In thermal leptogenesis, when $K\gg 1$, the Yukawa interactions are strong to bring the RHNs into thermal equilibrium. The premise that the RHNs are in or produced in thermal equilibrium makes all the differences. Although Yukawa interactions are strong in our considered scenario, the RHNs decay also fast. In the meantime, the RHNs are produced in low number density, indicating that they cannot interact quickly enough to thermalize. Additionally, this non-thermal limit contains the $K$ range that is preferred by neutrino oscillation experiments and has a relatively large final efficiency - larger than the thermal one with the same $K$. Contrary to the thermal leptogenesis situation with $M_1^{}>T$ to be Boltzmann suppressed and neglected, this {\it strongly non-thermal RHNs} limit occupies a large parameter space and is quite relevant.
\item The lower limit for $M_1^{}$ is examined in both unflavored and two-flavored treatments. In the unflavored case, the lowest value of $M_1^{}$ generating a large enough baryon asymmetry is $10^7$ GeV. Flavor effects further lower this value to $6\times 10^6$ GeV. This bound is $M_1^{} \geq 2\times 10^7$ GeV in the oscillation preferred range. The reheating temperature can be lowered greatly, and a conservative estimate gives $T_\mathrm{RH}^{} \geq 10^3$ GeV.
\end{itemize}

In this section, we present a general investigation of non-thermal leptogenesis from inflaton decay with least model-dependence. Partially the same subject has been investigated in Refs.~\cite{Chung:1998rq,Hahn-Woernle:2008tsk}. Non-thermal leptogenesis compared to thermal and preheating cases is presented in Ref.~\cite{Chung:1998rq} with a focus on scenarios that the analytical estimates apply. Ref.~\cite{Hahn-Woernle:2008tsk} mainly focuses on the case that $M_1^{} \sim T_\mathrm{RH}^{}$. Discussions in this section complete the studies in Refs.~\cite{Chung:1998rq,Hahn-Woernle:2008tsk}. We present the first systematic analysis of non-thermal leptogenesis to our knowledge. We give a general classification that is closely related to thermal leptogenesis and find four characteristic limits. The dynamics of the limits are examined both analytically and numerically. We also give the $\kappa_f^{}$-$K$ relation in comparison with thermal leptogenesis and discuss the lower limit of $M_1^{}$ and $T_\mathrm{RH}^{}$. We find the {\it strongly non-thermal RHNs} scenario particularly interesting. It opens up parameter space greatly in the range of $K$ that is preferred by the neutrino oscillations. Additionally, we investigate the effects of flavor and find the lower limit of $M_1^{}$ can be further lowered to $6\times 10^6$ GeV in the range $0.1\leq K \leq3$.

\section{Connecting to inflation through LNSB}\label{sec:inflation}

The discussion in the previous section is largely general, with the only assumption that the inflaton couples directly to the RHNs. It effectively puts no constraints on inflation except allowing a low reheating temperature. The inflaton mass can be any value satisfying the kinetic bound. To find a fully consistent picture with both the baryon asymmetry and inflation observations, one has to identify the inflaton and its potential to check the inflation dynamics. We consider the simplest scenario for inflation dynamics: a scalar field rolling down toward its potential minimum drives inflation. Such single-field slow-roll models are characterized by the inflaton potential, from which the observables can be derived. The slow-roll dynamics are described by the slow-roll parameters, which are calculated from the potential,
\begin{align}
\epsilon_V^{} = \frac{M_\mathrm{pl}^2}{2} \left(\frac{V_{,\phi}^{}}{V}\right)^2\;,\quad
\eta_V^{} = M_\mathrm{pl}^2 \frac{V_{,\phi\phi}}{V}\;,
\end{align}
where $M_\mathrm{pl}^{}$ is the reduced Planck mass, $V_{,\phi}^{} \equiv \mathrm{d} V/ \mathrm{d} \phi$. 

Inflation models are tested by the CMB anisotropy measurements, i.e., the spectral tilt $n_s^{}$ of the scalar power spectrum and the tensor-to-scalar ratio $r$, which can be expressed in terms of the slow-roll parameters as
\begin{align}
n_s^{} = 1- 6 \epsilon_V^{} + 2 \eta_V^{}\;,\quad
r = 16\epsilon_V^{}\;.
\end{align}
The Planck 2018 release determines the $n_s^{}$ and $r$ at the pivot scale $k=0.002~\mathrm{Mpc}^{-1}$ to be~\cite{Planck:2018jri} 
\begin{align}
n_s^{}=0.9649\pm 0.0126,\quad r<0.056,~(95\% ~\mathrm{C.L.})\;. \label{eq:inflation_obs}
\end{align}
The inflation scale gets constrained by the scalar power spectrum amplitude $A_s^{}$ and $r$ from
\begin{align}
V_*^{}=\frac{3\pi^2}{2} A_s^{} r M_\mathrm{pl}^4 < \left( 1.6\times 10^{16} ~\mathrm{GeV} \right)^4\;,\label{eq:COBEnorm}
\end{align}
where $*$ means it is evaluated at the horizon crossing, and the last inequality is obtained with current constrained $A_s^{}$ and $r$ at a $95\%$ confidence level~\cite{Planck:2018jri}. 

We consider the most general type-I seesaw extension of the SM with three RHNs $N_i^{}$ (for $i=1,2,3$) and one complex scalar $\sigma$. The relevant Lagrangian adding to the SM Lagrangian is 
\begin{align}
\mathcal{L} \supset \overline{N} i \slashed{\partial} N + (\partial^\mu_{} \sigma^\dagger)(\partial_\mu^{} \sigma)-\bar{L} Y_\nu^{}\widetilde{H} N - y_N^{} \sigma \overline{N^c_{}} N - V(H,\sigma) + \mathrm{h.c.} \;, \label{eq:LNSB}
\end{align}
where $L$ is the SM lepton doublet, $H$ is the Higgs doublet and $\widetilde{H}=i\sigma_2 H^*_{}$, $Y_\nu^{}$ is the Yukawa coupling matrix of the light and the heavy neutrinos, $y_N^{}$ is the coupling of the singlet scalar to the RHNs. We assign the lepton number $\mathrm{L}= 1$ for $N_i^{}$ and $\mathrm{L}= -2$ for $\sigma$. This framework features lepton number conserving and the RHNs become massive once $\sigma$ gets a vev. The lepton number spontaneously breaking (LNSB) is known as the singlet Majoron model~\cite{Chikashige:1980ui,Chikashige:1980qk}. We leave aside the dynamics of LNSB but focus on non-thermal leptogenesis and its connection to inflation. One of the degrees of freedom will be identified as the inflaton $\phi$, with a potential to be specified in the following.

\subsection{$\phi=\mathrm{Re}(\sigma)$: Coleman-Weinberg potential}\label{sec:cwi}

\begin{figure}[t!]
\centering
\includegraphics[width=.5\textwidth]{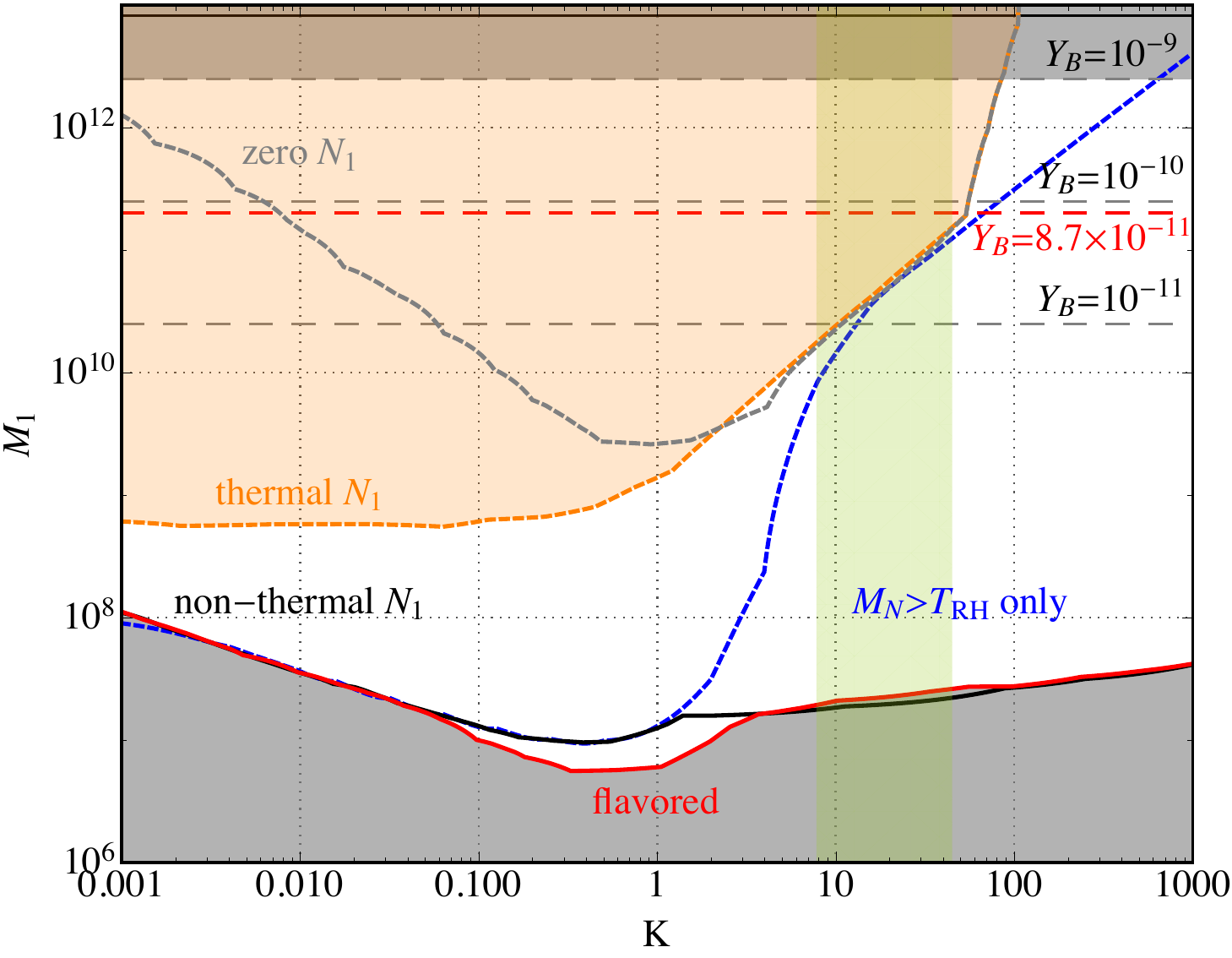}
\caption{The projection of the parameter space in the $M_1^{}$ - $K$ plane when the inflaton $\phi=\mathrm{Re}(\sigma)$ takes a Coleman-Weinberg potential. Also shown are the values of $Y_\mathrm{B}^{}$ with the red dashed line marking the observed value of $Y_\mathrm{B}^{}$.  The viable parameter space generating a large enough baryon asymmetry and compatible with inflation observation is the region above the red dashed line and below the solid black line at the top.}
\label{fig:mk_cwi}
\end{figure}

In the case that the inflaton corresponds to the real part of $\sigma$, 
we take its potential to be the Coleman-Weinberg potential~\cite{Coleman:1973jx,Rehman:2008qs,Okada:2014lxa}, i.e.,
\begin{align}
V(\phi) = \displaystyle A \phi^4 \left[ \mathrm{ln} \left(\frac{\phi}{v_\phi^{}} \right) -\frac{1}{4}\right] + \frac{1}{4} A v_\phi^4 \;,
\end{align}
where $v_\phi^{}=\langle \phi \rangle$, $V(0) =\frac{1}{4} A v_\phi^4$ is the vacuum energy at origin. The inflaton mass is $M_\phi=2\sqrt{A} v_\phi^{}$.

We take the benchmark point $A=2.41\times 10^{-14}, v_\phi^{}=22.1 M_\mathrm{pl}$ from Ref.~\cite{Okada:2014lxa} where $M_\mathrm{pl}^{}=m_\mathrm{pl}/\sqrt{8\pi}$ is the reduced Planck mass. This point gives $n_s^{}=0.964, r=0.036$ compatible with the Planck 2018 constraints~\cite{Planck:2018jri} and $M_\phi^{} = 1.65\times 10^{13}$ GeV. The value of $M_\phi^{}$ also implies an upper limit on the reheating temperature $T_\mathrm{RH}^{} < 1.21\times 10^{16}$ GeV considering the perturbativity bound of $y_N^{}$. From Eq.~(\ref{eq:LNSB}), the RHN mass $M_N^{}=y_N^{} v_\phi^{}$ can vary from the lower bound $10^7$ GeV to roughly $M_\phi^{}/2 \simeq10^{13}$ GeV, corresponding to different values of the coupling $y_N^{}$. It further allows us to relate the reheating temperature to the RHN mass directly, $T_\mathrm{RH}^{}\simeq 1.8\times 10^{-5} M_N^{}$. 

Regarding leptogenesis, the system has only two free parameters: $M_1^{}$ and $K$. The $T_\mathrm{RH}^{}$-$M_N^{}$ relation implies $T_*^{}>T_\mathrm{RH}^{}$. As a result, it is {\it instantaneous reheating} for the small $K$ end and {\it strongly non-thermal RHNs} for the large $K$ end. The final baryon asymmetry can be evaluated in both limits through Eq.~(\ref{eq:YBdecay}). We show the results on the $M_1^{}$ - $K$ plane in Fig.~\ref{fig:mk_cwi}.
We find that although inflation puts no constraints on the RHN mass, the requirement that generating a large enough baryon asymmetry in connection to this inflation model constrains $2.0\times 10^{11}\leq M_1^{} (\mathrm{GeV}) \leq 8.3\times 10^{12}$.

\subsection{$\phi=\mathrm{Im}(\sigma)$: Natural inflation potential}\label{sec:ni}

\begin{figure}[t!]
\centering
\includegraphics[width=.5\textwidth]{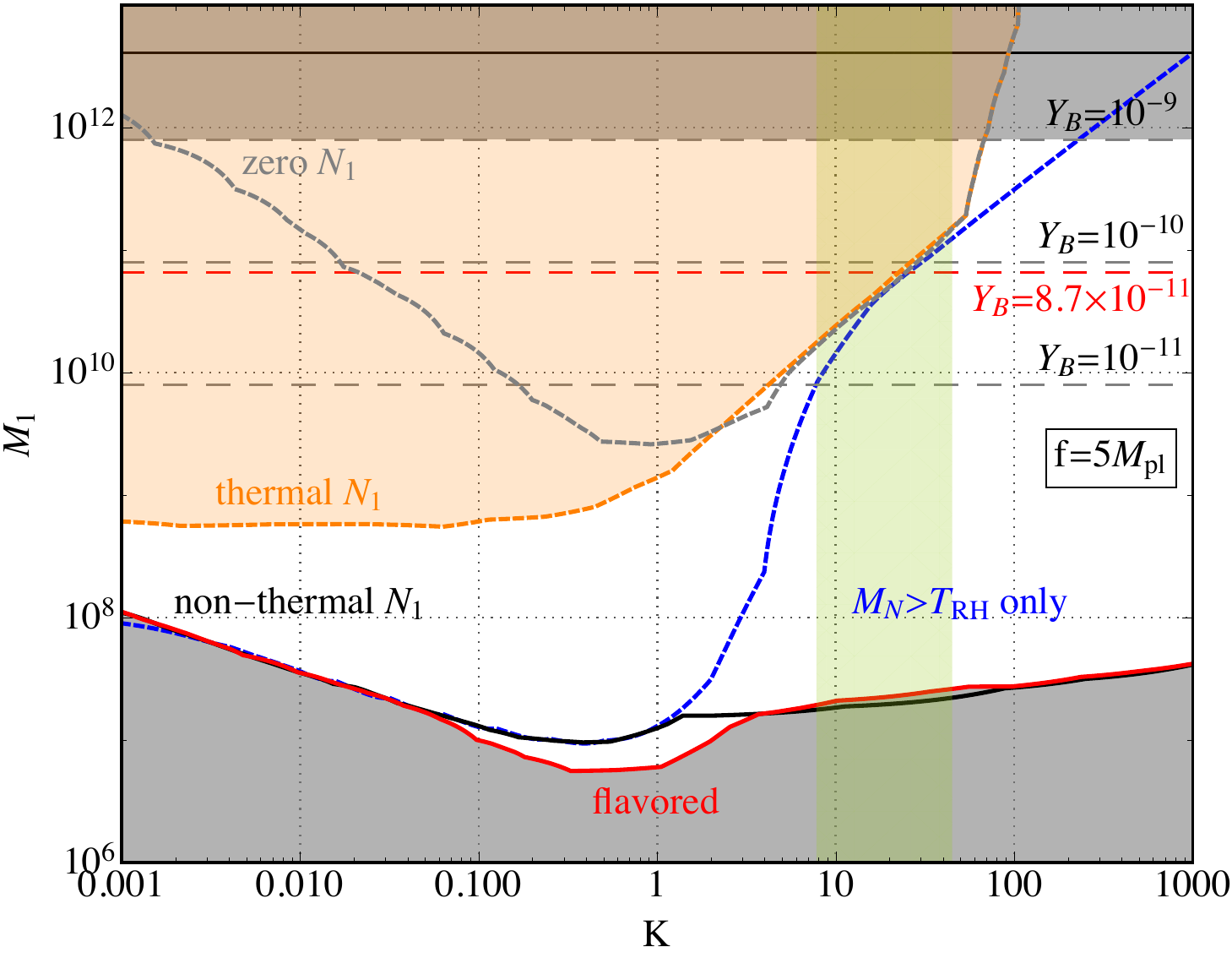}
\caption{The projection of the parameter space in the $M_1^{}$ - $K$ plane when the inflaton $\phi=\mathrm{Im}(\sigma)$ takes a natural inflation potential and $f=5 M_\mathrm{pl}$. Also shown are the values of $Y_\mathrm{B}^{}$ with the red dashed line marking the observed value of $Y_\mathrm{B}^{}$.  The viable parameter space generating a large enough baryon asymmetry and compatible with inflation observation is the region above the red dashed line and below the solid black line at the top.}
\label{fig:mk_ni}
\end{figure}

We consider the possibility that the pseudoscalar associated with lepton number spontaneously breaking is the inflaton. A natural choice of its potential is of the natural inflation form~\cite{Freese:1990rb,Adams:1992bn} 
\begin{align}
V(\phi)= \Lambda \left(1+\cos \phi/f \right)\;,\label{eq:NIpotential}
\end{align}
where $\Lambda$ is the inflation scale and $f$ is the lepton-number breaking scale. This potential is in close analog with the effective axion potential after chiral symmetry breaking. It possesses a quasi-shift symmetry that $\phi \rightarrow 2\pi f\phi$ leaves the potential invariant.

From Eq.~(\ref{eq:COBEnorm}), one obtains roughly $\Lambda \leq 10^{16} ~\mathrm{GeV}$ for the Majoron potential in Eq.~(\ref{eq:NIpotential}).
Regarding the slow-roll dynamics, this potential involves only one parameter $f$ to be constrained by $n_s^{}$ and $r$. The natural inflation model is at odds with the Planck 2018 results in its original form. However, if other effects are taken into consideration, there will be no problem for this model predictions compatible with observation. Several studies provide different methods, among which we list a few: in quadratic gravity~\cite{Salvio:2019wcp,Salvio:2020axm}, allowing a nonminimal coupling to the Ricci scalar together with a non-canonical kinetic term~\cite{Simeon:2020lkd}, with dissipative effects~\cite{Reyimuaji:2020bkm}, only with a nonminimal coupling to Ricci scalar~\cite{Reyimuaji:2020goi}, with a coupling to the trace of energy-momentum tensor~\cite{Zhang:2021ppy}, in $f(R,T)$ gravity~\cite{Chen:2022dyq}.
Different methods lead to a different range of $f$ that is compatible with observation. Generally, inflation observation favors a super-Planckian value of $f$ as in the original model and also in Refs.~\cite{Salvio:2019wcp,Salvio:2020axm,Reyimuaji:2020bkm,Reyimuaji:2020goi,Zhang:2021ppy,Chen:2022dyq,Salvio:2021lka}. For methods involving modified gravity, usually the Einstein frame potential will be modified. But the modifications are small - constrained by the success of General Relativity and the inflation observation - leading to negligible modifications for investigating the inflaton mass.

One can determine the inflaton mass from the potential in Eq.~(\ref{eq:NIpotential}) as
\begin{align}
M_\phi^2= \frac{\mathrm{d}^2 V}{\mathrm{d}\phi^2}|_\mathrm{min}^{} = \frac{\Lambda^4}{f^2}\;, \label{eq:mPhivf}
\end{align}
which is related to $f$ through the inflation energy scale $\Lambda$. As $\Lambda$ is constrained by observation to be $\Lambda \leq 10^{16}$ GeV, a super-Planckian $f\geq 10^{19}$ GeV requires $M_\phi^{} \leq 10^{13}$ GeV. For the fixed super-Planckian value of $f$, a lower inflation energy scale means a lower inflaton mass, a lower RHN mass, and a lower $T_\mathrm{RH}^{}$. We fix $M_\phi^{}=10^{13}$ GeV for definiteness. Since $\langle \sigma \rangle = f$, the RHN mass is $M_N^{}=y_N^{} f$, which is translated into $T_\mathrm{RH}^{}\simeq 6\times 10^{-5} M_N^{}$ for $f=5 M_\mathrm{pl}^{}$. Similar to the previous case, we still have in general $T_*^{} > T_\mathrm{RH}^{}$ and the final baryon asymmetry can be evaluated from Eq.~(\ref{eq:YBdecay}).
We plot $f=5 M_\mathrm{pl}^{}$ results in Fig.~\ref{fig:mk_ni}. We find a slighter larger range of $M_1^{}$ that generates a large enough $Y_\mathrm{B}^{}$ and agrees with inflation:  $6.6\times 10^{10} \leq M_1^{}(\mathrm{GeV}) \leq 4.1 \times 10^{12}$, corresponding $T_\mathrm{RH}^{}$ in $[4.1\times 10^6, 2.6\times 10^8]$ GeV.

Alternatively, with the method introduced in Ref.~\cite{Simeon:2020lkd}, $f$ can be arbitrarily low with no additional requirement for the inflaton mass. $y_N^{}$ can be $\mathcal{O}(1)$ in this case. Meanwhile, the reheating temperature is generally larger than $M_1^{}$, allowing the RHNs to be thermalized when $K\gg 1$ and the $M_1^{}$ - $K$ relation takes the form of the blue-dashed line in Fig.~\ref{fig:mk_ni}.

\section{Conclusions}\label{sec:conclusion}

This work considers connecting inflation and neutrino physics through non-thermal leptogenesis from inflaton decay. Assuming a direct inflaton-RHN coupling, we present the first systematic investigation on non-thermal leptogenesis. We give a general classification of the parameter space and find four characteristic limits with working conditions to identify them (see Fig.~\ref{fig:ntClass}). Three of the four limits are truly non-thermal, with a final efficiency larger than that of thermal leptogenesis (see Fig.~\ref{fig:kk_compare}). Two analytic estimates (Eqs.~(\ref{eq:YBNdom}) and (\ref{eq:YBdecay}) work for these three limits. We find the lower limit of $M_1^{}$ in the oscillation-preferred region is $2\times 10^7$ GeV. When flavor effects are included, it can be lowered to $6\times 10^6$ GeV. The reheating temperature can be as low as $10^3$ GeV. 

Then we assume a simple and natural connection of inflaton with the RHNs through LNSB by identifying either the real or the imaginary part of the complex field $\sigma$ responsible for LNSB to be the inflaton. We consider the inflaton possesses either a Coleman-Weinberg potential (for the real part of $\sigma$) or a natural inflation potential (for the imaginary part of $\sigma$) and find viable parameter space satisfying constraints from inflation, BAU, and neutrino data, which is more constrained than considering non-thermal leptogenesis alone (see Figs.~\ref{fig:mk_cwi} and \ref{fig:mk_ni}). LNSB suppresses the inflaton-RHN coupling greatly, leading to a low $T_\mathrm{RH}^{}$ and resulting non-thermal leptogenesis belongs to intantaneous RHN decay category in both models, in the {\it strongly non-thermal RHNs} limit for large $K$ and in the {\it intantaneous reheating} limit for small $K$. 

We find the {\it strongly non-thermal RHNs} scenario particularly interesting. Although Yukawa interactions are strong, the RHNs produced in low number density decay very fast to avoid entering thermal equilibrium. It features a low reheating temperature and occupies a large parameter space, including the oscillation-preferred $K$ range.

Non-thermal leptogenesis from inflaton decay offers a testable framework for the early Universe. This framework contains well-motivated ingredients (inflation, type-I seesaw, leptogenesis) and offers simple and elegant solutions to important questions in cosmology and particle physics. It can be further constrained and tested with upcoming cosmological and neutrino data. The two examples we show with LNSB connection are only illustrative. The model-independent investigation of non-thermal leptogenesis should be useful in exploring this direction in the future.

\section*{Acknowledgements}
The author is indebted to Prof. Shun Zhou for many valuable discussions. This work is supported by the National Natural Science Foundation of China under grant Nos.~11835013 and 12305116, the Start-up Funds for Young Talents of Hebei University (No. 521100223012).

\bibliographystyle{JHEP}
\bibliography{bibliography}

\providecommand{\href}[2]{#2}\begingroup\raggedright\begin{thebibliography}{10}

\bibitem{Esteban:2020cvm}
I.~Esteban, M.C.~Gonzalez-Garcia, M.~Maltoni, T.~Schwetz and A.~Zhou,
  \emph{{The fate of hints: updated global analysis of three-flavor neutrino
  oscillations}}, \href{https://doi.org/10.1007/JHEP09(2020)178}{\emph{JHEP}
  {\bfseries 09} (2020) 178}
  [\href{https://arxiv.org/abs/2007.14792}{{\ttfamily 2007.14792}}].

\bibitem{Planck:2018vyg}
{\scshape Planck} collaboration, \emph{{Planck 2018 results. VI. Cosmological
  parameters}},
  \href{https://doi.org/10.1051/0004-6361/201833910}{\emph{Astron. Astrophys.}
  {\bfseries 641} (2020) A6}
  [\href{https://arxiv.org/abs/1807.06209}{{\ttfamily 1807.06209}}].

\bibitem{Giudice:2003jh}
G.F.~Giudice, A.~Notari, M.~Raidal, A.~Riotto and A.~Strumia, \emph{{Towards a
  complete theory of thermal leptogenesis in the SM and MSSM}},
  \href{https://doi.org/10.1016/j.nuclphysb.2004.02.019}{\emph{Nucl. Phys. B}
  {\bfseries 685} (2004) 89}
  [\href{https://arxiv.org/abs/hep-ph/0310123}{{\ttfamily hep-ph/0310123}}].

\bibitem{Buchmuller:2004nz}
W.~Buchmuller, P.~Di~Bari and M.~Plumacher, \emph{{Leptogenesis for
  pedestrians}}, \href{https://doi.org/10.1016/j.aop.2004.02.003}{\emph{Annals
  Phys.} {\bfseries 315} (2005) 305}
  [\href{https://arxiv.org/abs/hep-ph/0401240}{{\ttfamily hep-ph/0401240}}].

\bibitem{Davidson:2008bu}
S.~Davidson, E.~Nardi and Y.~Nir, \emph{{Leptogenesis}},
  \href{https://doi.org/10.1016/j.physrep.2008.06.002}{\emph{Phys. Rept.}
  {\bfseries 466} (2008) 105}
  [\href{https://arxiv.org/abs/0802.2962}{{\ttfamily 0802.2962}}].

\bibitem{Xing:2020ald}
Z.-z.~Xing and Z.-h.~Zhao, \emph{{The minimal seesaw and leptogenesis models}},
  \href{https://doi.org/10.1088/1361-6633/abf086}{\emph{Rept. Prog. Phys.}
  {\bfseries 84} (2021) 066201}
  [\href{https://arxiv.org/abs/2008.12090}{{\ttfamily 2008.12090}}].

\bibitem{Bodeker:2020ghk}
D.~Bodeker and W.~Buchmuller, \emph{{Baryogenesis from the weak scale to the
  grand unification scale}},
  \href{https://doi.org/10.1103/RevModPhys.93.035004}{\emph{Rev. Mod. Phys.}
  {\bfseries 93} (2021) 035004}
  [\href{https://arxiv.org/abs/2009.07294}{{\ttfamily 2009.07294}}].

\bibitem{Xing:2011zza}
Z.-z.~Xing and S.~Zhou, \emph{{Neutrinos in particle physics, astronomy and
  cosmology}}, Springer-Verlag, Berlin Heidelberg (2011).

\bibitem{Fukugita:1986hr}
M.~Fukugita and T.~Yanagida, \emph{{Baryogenesis Without Grand Unification}},
  \href{https://doi.org/10.1016/0370-2693(86)91126-3}{\emph{Phys. Lett. B}
  {\bfseries 174} (1986) 45}.

\bibitem{Branco:2001pq}
G.C.~Branco, T.~Morozumi, B.M.~Nobre and M.N.~Rebelo, \emph{{A Bridge between
  CP violation at low-energies and leptogenesis}},
  \href{https://doi.org/10.1016/S0550-3213(01)00425-4}{\emph{Nucl. Phys. B}
  {\bfseries 617} (2001) 475}
  [\href{https://arxiv.org/abs/hep-ph/0107164}{{\ttfamily hep-ph/0107164}}].

\bibitem{Branco:2002xf}
G.C.~Branco, R.~Gonzalez~Felipe, F.R.~Joaquim, I.~Masina, M.N.~Rebelo and
  C.A.~Savoy, \emph{{Minimal scenarios for leptogenesis and CP violation}},
  \href{https://doi.org/10.1103/PhysRevD.67.073025}{\emph{Phys. Rev. D}
  {\bfseries 67} (2003) 073025}
  [\href{https://arxiv.org/abs/hep-ph/0211001}{{\ttfamily hep-ph/0211001}}].

\bibitem{Pascoli:2006ie}
S.~Pascoli, S.T.~Petcov and A.~Riotto, \emph{{Connecting low energy leptonic
  CP-violation to leptogenesis}},
  \href{https://doi.org/10.1103/PhysRevD.75.083511}{\emph{Phys. Rev. D}
  {\bfseries 75} (2007) 083511}
  [\href{https://arxiv.org/abs/hep-ph/0609125}{{\ttfamily hep-ph/0609125}}].

\bibitem{Pascoli:2006ci}
S.~Pascoli, S.T.~Petcov and A.~Riotto, \emph{{Leptogenesis and Low Energy CP
  Violation in Neutrino Physics}},
  \href{https://doi.org/10.1016/j.nuclphysb.2007.02.019}{\emph{Nucl. Phys. B}
  {\bfseries 774} (2007) 1}
  [\href{https://arxiv.org/abs/hep-ph/0611338}{{\ttfamily hep-ph/0611338}}].

\bibitem{Moffat:2018smo}
K.~Moffat, S.~Pascoli, S.T.~Petcov and J.~Turner, \emph{{Leptogenesis from Low
  Energy $CP$ Violation}},
  \href{https://doi.org/10.1007/JHEP03(2019)034}{\emph{JHEP} {\bfseries 03}
  (2019) 034} [\href{https://arxiv.org/abs/1809.08251}{{\ttfamily
  1809.08251}}].

\bibitem{Xing:2020erm}
Z.-z.~Xing and D.~Zhang, \emph{{A direct link between unflavored leptogenesis
  and low-energy CP violation via the one-loop quantum corrections}},
  \href{https://doi.org/10.1007/JHEP04(2020)179}{\emph{JHEP} {\bfseries 04}
  (2020) 179} [\href{https://arxiv.org/abs/2003.00480}{{\ttfamily
  2003.00480}}].

\bibitem{Xing:2020ghj}
Z.-z.~Xing and D.~Zhang, \emph{{Bridging resonant leptogenesis and low-energy
  CP violation with an RGE-modified seesaw relation}},
  \href{https://doi.org/10.1016/j.physletb.2020.135397}{\emph{Phys. Lett. B}
  {\bfseries 804} (2020) 135397}
  [\href{https://arxiv.org/abs/2003.06312}{{\ttfamily 2003.06312}}].

\bibitem{Zhang:2020lir}
X.~Zhang, J.-H.~Yu and B.-Q.~Ma, \emph{{Leptogenesis from low-energy CP
  violation in minimal left-right symmetric model}},
  \href{https://doi.org/10.1016/j.nuclphysb.2022.115670}{\emph{Nucl. Phys. B}
  {\bfseries 976} (2022) 115670}
  [\href{https://arxiv.org/abs/2008.06433}{{\ttfamily 2008.06433}}].

\bibitem{Bi:2003yr}
X.-J.~Bi, P.-h.~Gu, X.-l.~Wang and X.-m.~Zhang, \emph{{Thermal leptogenesis in
  a model with mass varying neutrinos}},
  \href{https://doi.org/10.1103/PhysRevD.69.113007}{\emph{Phys. Rev. D}
  {\bfseries 69} (2004) 113007}
  [\href{https://arxiv.org/abs/hep-ph/0311022}{{\ttfamily hep-ph/0311022}}].

\bibitem{Xing:2006ms}
Z.-z.~Xing and S.~Zhou, \emph{{Tri-bimaximal Neutrino Mixing and
  Flavor-dependent Resonant Leptogenesis}},
  \href{https://doi.org/10.1016/j.physletb.2007.08.009}{\emph{Phys. Lett. B}
  {\bfseries 653} (2007) 278}
  [\href{https://arxiv.org/abs/hep-ph/0607302}{{\ttfamily hep-ph/0607302}}].

\bibitem{Guo:2006qa}
W.-l.~Guo, Z.-z.~Xing and S.~Zhou, \emph{{Neutrino Masses, Lepton Flavor Mixing
  and Leptogenesis in the Minimal Seesaw Model}},
  \href{https://doi.org/10.1142/S0218301307004898}{\emph{Int. J. Mod. Phys. E}
  {\bfseries 16} (2007) 1}
  [\href{https://arxiv.org/abs/hep-ph/0612033}{{\ttfamily hep-ph/0612033}}].

\bibitem{Gu:2010xc}
P.-H.~Gu and U.~Sarkar, \emph{{Leptogenesis with Linear, Inverse or Double
  Seesaw}}, \href{https://doi.org/10.1016/j.physletb.2010.09.062}{\emph{Phys.
  Lett. B} {\bfseries 694} (2011) 226}
  [\href{https://arxiv.org/abs/1007.2323}{{\ttfamily 1007.2323}}].

\bibitem{Gehrlein:2015dxa}
J.~Gehrlein, S.T.~Petcov, M.~Spinrath and X.~Zhang, \emph{{Leptogenesis in an
  SU(5) $\times$ A$_5$ Golden Ratio Flavour Model}},
  \href{https://doi.org/10.1016/j.nuclphysb.2015.04.019}{\emph{Nucl. Phys. B}
  {\bfseries 896} (2015) 311}
  [\href{https://arxiv.org/abs/1502.00110}{{\ttfamily 1502.00110}}].

\bibitem{Gehrlein:2015dza}
J.~Gehrlein, S.T.~Petcov, M.~Spinrath and X.~Zhang, \emph{{Leptogenesis in an
  SU(5) x A5 Golden Ratio Flavour Model: Addendum}},
  \href{https://arxiv.org/abs/1508.07930}{{\ttfamily 1508.07930}}.

\bibitem{Sarma:2022qka}
L.~Sarma, P.K.~Paul and M.K.~Das, \emph{{Connecting dark matter, baryogenesis
  and neutrinoless double beta decay in a A4\ensuremath{\otimes}Z8-based
  \ensuremath{\nu}2HDM}},
  \href{https://doi.org/10.1142/S0217751X22501573}{\emph{Int. J. Mod. Phys. A}
  {\bfseries 37} (2022) 2250157}
  [\href{https://arxiv.org/abs/2208.14764}{{\ttfamily 2208.14764}}].

\bibitem{Mahapatra:2023dbr}
S.~Mahapatra, P.K.~Paul, N.~Sahu and P.~Shukla, \emph{{Cogenesis of matter and
  dark matter from triplet fermion seesaw}},
  \href{https://arxiv.org/abs/2305.11138}{{\ttfamily 2305.11138}}.

\bibitem{Lazarides:1990huy}
G.~Lazarides and Q.~Shafi, \emph{{Origin of matter in the inflationary
  cosmology}}, \href{https://doi.org/10.1016/0370-2693(91)91090-I}{\emph{Phys.
  Lett. B} {\bfseries 258} (1991) 305}.

\bibitem{Guth:1980zm}
A.H.~Guth, \emph{{The Inflationary Universe: A Possible Solution to the Horizon
  and Flatness Problems}},
  \href{https://doi.org/10.1103/PhysRevD.23.347}{\emph{Phys. Rev. D} {\bfseries
  23} (1981) 347}.

\bibitem{Bassett:2005xm}
B.A.~Bassett, S.~Tsujikawa and D.~Wands, \emph{{Inflation dynamics and
  reheating}}, \href{https://doi.org/10.1103/RevModPhys.78.537}{\emph{Rev. Mod.
  Phys.} {\bfseries 78} (2006) 537}
  [\href{https://arxiv.org/abs/astro-ph/0507632}{{\ttfamily
  astro-ph/0507632}}].

\bibitem{Baumann:2009ds}
D.~Baumann, \emph{{Inflation}},  in \emph{{Theoretical Advanced Study Institute
  in Elementary Particle Physics}: {Physics of the Large and the Small}},
  pp.~523--686, 2011, \href{https://doi.org/10.1142/9789814327183_0010}{DOI}
  [\href{https://arxiv.org/abs/0907.5424}{{\ttfamily 0907.5424}}].

\bibitem{Kumekawa:1994gx}
K.~Kumekawa, T.~Moroi and T.~Yanagida, \emph{{Flat potential for inflaton with
  a discrete R invariance in supergravity}},
  \href{https://doi.org/10.1143/PTP.92.437}{\emph{Prog. Theor. Phys.}
  {\bfseries 92} (1994) 437}
  [\href{https://arxiv.org/abs/hep-ph/9405337}{{\ttfamily hep-ph/9405337}}].

\bibitem{Asaka:1999yd}
T.~Asaka, K.~Hamaguchi, M.~Kawasaki and T.~Yanagida, \emph{{Leptogenesis in
  inflaton decay}},
  \href{https://doi.org/10.1016/S0370-2693(99)01020-5}{\emph{Phys. Lett. B}
  {\bfseries 464} (1999) 12}
  [\href{https://arxiv.org/abs/hep-ph/9906366}{{\ttfamily hep-ph/9906366}}].

\bibitem{Lazarides:1999dm}
G.~Lazarides, \emph{{Leptogenesis in supersymmetric hybrid inflation}},
  {\emph{Springer Tracts Mod. Phys.} {\bfseries 163} (2000) 227}
  [\href{https://arxiv.org/abs/hep-ph/9904428}{{\ttfamily hep-ph/9904428}}].

\bibitem{Jeannerot:2001qu}
R.~Jeannerot, S.~Khalil and G.~Lazarides, \emph{{Leptogenesis in smooth hybrid
  inflation}}, \href{https://doi.org/10.1016/S0370-2693(01)00429-4}{\emph{Phys.
  Lett. B} {\bfseries 506} (2001) 344}
  [\href{https://arxiv.org/abs/hep-ph/0103229}{{\ttfamily hep-ph/0103229}}].

\bibitem{Senoguz:2003hc}
V.N.~Senoguz and Q.~Shafi, \emph{{GUT scale inflation, nonthermal leptogenesis,
  and atmospheric neutrino oscillations}},
  \href{https://doi.org/10.1016/j.physletb.2003.12.020}{\emph{Phys. Lett. B}
  {\bfseries 582} (2004) 6}
  [\href{https://arxiv.org/abs/hep-ph/0309134}{{\ttfamily hep-ph/0309134}}].

\bibitem{Dent:2003dn}
T.~Dent, G.~Lazarides and R.~Ruiz~de Austri, \emph{{Leptogenesis through direct
  inflaton decay to light particles}},
  \href{https://doi.org/10.1103/PhysRevD.69.075012}{\emph{Phys. Rev. D}
  {\bfseries 69} (2004) 075012}
  [\href{https://arxiv.org/abs/hep-ph/0312033}{{\ttfamily hep-ph/0312033}}].

\bibitem{Dent:2005gx}
T.~Dent, G.~Lazarides and R.~Ruiz~de Austri, \emph{{Non-thermal leptogenesis
  via direct inflaton decay without SU(2)(L) triplets}},
  \href{https://doi.org/10.1103/PhysRevD.72.043502}{\emph{Phys. Rev. D}
  {\bfseries 72} (2005) 043502}
  [\href{https://arxiv.org/abs/hep-ph/0503235}{{\ttfamily hep-ph/0503235}}].

\bibitem{Endo:2006nj}
M.~Endo, F.~Takahashi and T.T.~Yanagida, \emph{{Spontaneous Non-thermal
  Leptogenesis in High-scale Inflation Models}},
  \href{https://doi.org/10.1103/PhysRevD.74.123523}{\emph{Phys. Rev. D}
  {\bfseries 74} (2006) 123523}
  [\href{https://arxiv.org/abs/hep-ph/0611055}{{\ttfamily hep-ph/0611055}}].

\bibitem{Antusch:2010mv}
S.~Antusch, J.P.~Baumann, V.F.~Domcke and P.M.~Kostka, \emph{{Sneutrino Hybrid
  Inflation and Nonthermal Leptogenesis}},
  \href{https://doi.org/10.1088/1475-7516/2010/10/006}{\emph{JCAP} {\bfseries
  10} (2010) 006} [\href{https://arxiv.org/abs/1007.0708}{{\ttfamily
  1007.0708}}].

\bibitem{Khalil:2012nd}
S.~Khalil, Q.~Shafi and A.~Sil, \emph{{Smooth Hybrid Inflation and Non-Thermal
  Type II Leptogenesis}},
  \href{https://doi.org/10.1103/PhysRevD.86.073004}{\emph{Phys. Rev. D}
  {\bfseries 86} (2012) 073004}
  [\href{https://arxiv.org/abs/1208.0731}{{\ttfamily 1208.0731}}].

\bibitem{Murayama:1992ua}
H.~Murayama, H.~Suzuki, T.~Yanagida and J.~Yokoyama, \emph{{Chaotic inflation
  and baryogenesis by right-handed sneutrinos}},
  \href{https://doi.org/10.1103/PhysRevLett.70.1912}{\emph{Phys. Rev. Lett.}
  {\bfseries 70} (1993) 1912}.

\bibitem{Ellis:2003sq}
J.R.~Ellis, M.~Raidal and T.~Yanagida, \emph{{Sneutrino inflation in the light
  of WMAP: Reheating, leptogenesis and flavor violating lepton decays}},
  \href{https://doi.org/10.1016/j.physletb.2003.11.029}{\emph{Phys. Lett. B}
  {\bfseries 581} (2004) 9}
  [\href{https://arxiv.org/abs/hep-ph/0303242}{{\ttfamily hep-ph/0303242}}].

\bibitem{Pallis:2011ps}
C.~Pallis and N.~Toumbas, \emph{{Non-Minimal Sneutrino Inflation, Peccei-Quinn
  Phase Transition and non-Thermal Leptogenesis}},
  \href{https://doi.org/10.1088/1475-7516/2011/02/019}{\emph{JCAP} {\bfseries
  02} (2011) 019} [\href{https://arxiv.org/abs/1101.0325}{{\ttfamily
  1101.0325}}].

\bibitem{Pallis:2011gr}
C.~Pallis and N.~Toumbas, \emph{{Non-Minimal Higgs Inflation and non-Thermal
  Leptogenesis in A Supersymmetric Pati-Salam Model}},
  \href{https://doi.org/10.1088/1475-7516/2011/12/002}{\emph{JCAP} {\bfseries
  12} (2011) 002} [\href{https://arxiv.org/abs/1108.1771}{{\ttfamily
  1108.1771}}].

\bibitem{Pallis:2012iw}
C.~Pallis and Q.~Shafi, \emph{{Non-Minimal Chaotic Inflation, Peccei-Quinn
  Phase Transition and non-Thermal Leptogenesis}},
  \href{https://doi.org/10.1103/PhysRevD.86.023523}{\emph{Phys. Rev. D}
  {\bfseries 86} (2012) 023523}
  [\href{https://arxiv.org/abs/1204.0252}{{\ttfamily 1204.0252}}].

\bibitem{Antusch:2018zvu}
S.~Antusch and K.~Marschall, \emph{{Non-thermal Leptogenesis after Majoron
  Hilltop Inflation}},
  \href{https://doi.org/10.1088/1475-7516/2018/05/015}{\emph{JCAP} {\bfseries
  05} (2018) 015} [\href{https://arxiv.org/abs/1802.05647}{{\ttfamily
  1802.05647}}].

\bibitem{Panotopoulos:2021ttt}
G.~Panotopoulos, \emph{{Inflationary Universe with a Coleman-Weinberg potential
  meets non-thermal leptogenesis}},
  \href{https://doi.org/10.1016/j.astropartphys.2021.102559}{\emph{Astropart.
  Phys.} {\bfseries 128} (2021) 102559}.

\bibitem{SravanKumar:2018tgk}
K.~Sravan~Kumar and P.~Vargas~Moniz, \emph{{Conformal GUT inflation, proton
  lifetime and non-thermal leptogenesis}},
  \href{https://doi.org/10.1140/epjc/s10052-019-7449-1}{\emph{Eur. Phys. J. C}
  {\bfseries 79} (2019) 945}
  [\href{https://arxiv.org/abs/1806.09032}{{\ttfamily 1806.09032}}].

\bibitem{Fukuyama:2005us}
T.~Fukuyama, T.~Kikuchi and T.~Osaka, \emph{{Non-thermal leptogenesis and a
  prediction of inflaton mass in a supersymmetric SO(10) model}},
  \href{https://doi.org/10.1088/1475-7516/2005/06/005}{\emph{JCAP} {\bfseries
  06} (2005) 005} [\href{https://arxiv.org/abs/hep-ph/0503201}{{\ttfamily
  hep-ph/0503201}}].

\bibitem{Baer:2008eq}
H.~Baer and H.~Summy, \emph{{SO(10) SUSY GUTs, the gravitino problem,
  non-thermal leptogenesis and axino dark matter}},
  \href{https://doi.org/10.1016/j.physletb.2008.06.072}{\emph{Phys. Lett. B}
  {\bfseries 666} (2008) 5} [\href{https://arxiv.org/abs/0803.0510}{{\ttfamily
  0803.0510}}].

\bibitem{Fukuyama:2010hh}
T.~Fukuyama and N.~Okada, \emph{{Non-thermal Leptogenesis in a simple 5D SO(10)
  GUT}}, \href{https://doi.org/10.1088/1475-7516/2010/09/024}{\emph{JCAP}
  {\bfseries 09} (2010) 024} [\href{https://arxiv.org/abs/1003.2691}{{\ttfamily
  1003.2691}}].

\bibitem{Asaka:2002zu}
T.~Asaka, H.B.~Nielsen and Y.~Takanishi, \emph{{Nonthermal leptogenesis from
  the heavier Majorana neutrinos}},
  \href{https://doi.org/10.1016/S0550-3213(02)00934-3}{\emph{Nucl. Phys. B}
  {\bfseries 647} (2002) 252}
  [\href{https://arxiv.org/abs/hep-ph/0207023}{{\ttfamily hep-ph/0207023}}].

\bibitem{Allahverdi:2002gz}
R.~Allahverdi and A.~Mazumdar, \emph{{Nonthermal leptogenesis with almost
  degenerate superheavy neutrinos}},
  \href{https://doi.org/10.1103/PhysRevD.67.023509}{\emph{Phys. Rev. D}
  {\bfseries 67} (2003) 023509}
  [\href{https://arxiv.org/abs/hep-ph/0208268}{{\ttfamily hep-ph/0208268}}].

\bibitem{Panotopoulos:2006wj}
G.~Panotopoulos, \emph{{Non-thermal leptogenesis and baryon asymmetry in
  different neutrino mass models}},
  \href{https://doi.org/10.1016/j.physletb.2006.10.052}{\emph{Phys. Lett. B}
  {\bfseries 643} (2006) 279}
  [\href{https://arxiv.org/abs/hep-ph/0606127}{{\ttfamily hep-ph/0606127}}].

\bibitem{Senoguz:2007hu}
V.N.~Senoguz, \emph{{Non-thermal leptogenesis with strongly hierarchical right
  handed neutrinos}},
  \href{https://doi.org/10.1103/PhysRevD.76.013005}{\emph{Phys. Rev. D}
  {\bfseries 76} (2007) 013005}
  [\href{https://arxiv.org/abs/0704.3048}{{\ttfamily 0704.3048}}].

\bibitem{Ghoshal:2022fud}
A.~Ghoshal, D.~Nanda and A.K.~Saha, \emph{{CMB footprints of high scale
  non-thermal leptogenesis}},
  \href{https://arxiv.org/abs/2210.14176}{{\ttfamily 2210.14176}}.

\bibitem{Chung:1998rq}
D.J.H.~Chung, E.W.~Kolb and A.~Riotto, \emph{{Production of massive particles
  during reheating}},
  \href{https://doi.org/10.1103/PhysRevD.60.063504}{\emph{Phys. Rev. D}
  {\bfseries 60} (1999) 063504}
  [\href{https://arxiv.org/abs/hep-ph/9809453}{{\ttfamily hep-ph/9809453}}].

\bibitem{Hahn-Woernle:2008tsk}
F.~Hahn-Woernle and M.~Plumacher, \emph{{Effects of reheating on
  leptogenesis}},
  \href{https://doi.org/10.1016/j.nuclphysb.2008.07.032}{\emph{Nucl. Phys. B}
  {\bfseries 806} (2009) 68} [\href{https://arxiv.org/abs/0801.3972}{{\ttfamily
  0801.3972}}].

\bibitem{Giudice:1999fb}
G.F.~Giudice, M.~Peloso, A.~Riotto and I.~Tkachev, \emph{{Production of massive
  fermions at preheating and leptogenesis}},
  \href{https://doi.org/10.1088/1126-6708/1999/08/014}{\emph{JHEP} {\bfseries
  08} (1999) 014} [\href{https://arxiv.org/abs/hep-ph/9905242}{{\ttfamily
  hep-ph/9905242}}].

\bibitem{AristizabalSierra:2014uzi}
D.~Aristizabal~Sierra, M.~Tortola, J.W.F.~Valle and A.~Vicente,
  \emph{{Leptogenesis with a dynamical seesaw scale}},
  \href{https://doi.org/10.1088/1475-7516/2014/07/052}{\emph{JCAP} {\bfseries
  07} (2014) 052} [\href{https://arxiv.org/abs/1405.4706}{{\ttfamily
  1405.4706}}].

\bibitem{Davidson:2002qv}
S.~Davidson and A.~Ibarra, \emph{{A Lower bound on the right-handed neutrino
  mass from leptogenesis}},
  \href{https://doi.org/10.1016/S0370-2693(02)01735-5}{\emph{Phys. Lett. B}
  {\bfseries 535} (2002) 25}
  [\href{https://arxiv.org/abs/hep-ph/0202239}{{\ttfamily hep-ph/0202239}}].

\bibitem{Nardi:2006fx}
E.~Nardi, Y.~Nir, E.~Roulet and J.~Racker, \emph{{The Importance of flavor in
  leptogenesis}},
  \href{https://doi.org/10.1088/1126-6708/2006/01/164}{\emph{JHEP} {\bfseries
  01} (2006) 164} [\href{https://arxiv.org/abs/hep-ph/0601084}{{\ttfamily
  hep-ph/0601084}}].

\bibitem{Dev:2017trv}
P.S.B.~Dev, P.~Di~Bari, B.~Garbrecht, S.~Lavignac, P.~Millington and D.~Teresi,
  \emph{{Flavor effects in leptogenesis}},
  \href{https://doi.org/10.1142/S0217751X18420010}{\emph{Int. J. Mod. Phys. A}
  {\bfseries 33} (2018) 1842001}
  [\href{https://arxiv.org/abs/1711.02861}{{\ttfamily 1711.02861}}].

\bibitem{Planck:2018jri}
{\scshape Planck} collaboration, \emph{{Planck 2018 results. X. Constraints on
  inflation}}, \href{https://doi.org/10.1051/0004-6361/201833887}{\emph{Astron.
  Astrophys.} {\bfseries 641} (2020) A10}
  [\href{https://arxiv.org/abs/1807.06211}{{\ttfamily 1807.06211}}].

\bibitem{Chikashige:1980ui}
Y.~Chikashige, R.N.~Mohapatra and R.D.~Peccei, \emph{{Are There Real Goldstone
  Bosons Associated with Broken Lepton Number?}},
  \href{https://doi.org/10.1016/0370-2693(81)90011-3}{\emph{Phys. Lett. B}
  {\bfseries 98} (1981) 265}.

\bibitem{Chikashige:1980qk}
Y.~Chikashige, R.N.~Mohapatra and R.D.~Peccei, \emph{{Spontaneously Broken
  Lepton Number and Cosmological Constraints on the Neutrino Mass Spectrum}},
  \href{https://doi.org/10.1103/PhysRevLett.45.1926}{\emph{Phys. Rev. Lett.}
  {\bfseries 45} (1980) 1926}.

\bibitem{Coleman:1973jx}
S.R.~Coleman and E.J.~Weinberg, \emph{{Radiative Corrections as the Origin of
  Spontaneous Symmetry Breaking}},
  \href{https://doi.org/10.1103/PhysRevD.7.1888}{\emph{Phys. Rev. D} {\bfseries
  7} (1973) 1888}.

\bibitem{Rehman:2008qs}
M.U.~Rehman, Q.~Shafi and J.R.~Wickman, \emph{{GUT Inflation and Proton Decay
  after WMAP5}}, \href{https://doi.org/10.1103/PhysRevD.78.123516}{\emph{Phys.
  Rev. D} {\bfseries 78} (2008) 123516}
  [\href{https://arxiv.org/abs/0810.3625}{{\ttfamily 0810.3625}}].

\bibitem{Okada:2014lxa}
N.~Okada, V.N.~\c{S}eno\u{g}uz and Q.~Shafi, \emph{{The Observational Status of
  Simple Inflationary Models: an Update}},
  \href{https://doi.org/10.3906/fiz-1505-7}{\emph{Turk. J. Phys.} {\bfseries
  40} (2016) 150} [\href{https://arxiv.org/abs/1403.6403}{{\ttfamily
  1403.6403}}].

\bibitem{Freese:1990rb}
K.~Freese, J.A.~Frieman and A.V.~Olinto, \emph{{Natural inflation with pseudo -
  Nambu-Goldstone bosons}},
  \href{https://doi.org/10.1103/PhysRevLett.65.3233}{\emph{Phys. Rev. Lett.}
  {\bfseries 65} (1990) 3233}.

\bibitem{Adams:1992bn}
F.C.~Adams, J.R.~Bond, K.~Freese, J.A.~Frieman and A.V.~Olinto, \emph{{Natural
  inflation: Particle physics models, power law spectra for large scale
  structure, and constraints from COBE}},
  \href{https://doi.org/10.1103/PhysRevD.47.426}{\emph{Phys. Rev. D} {\bfseries
  47} (1993) 426} [\href{https://arxiv.org/abs/hep-ph/9207245}{{\ttfamily
  hep-ph/9207245}}].

\bibitem{Salvio:2019wcp}
A.~Salvio, \emph{{Quasi-Conformal Models and the Early Universe}},
  \href{https://doi.org/10.1140/epjc/s10052-019-7267-5}{\emph{Eur. Phys. J. C}
  {\bfseries 79} (2019) 750}
  [\href{https://arxiv.org/abs/1907.00983}{{\ttfamily 1907.00983}}].

\bibitem{Salvio:2020axm}
A.~Salvio, \emph{{Dimensional Transmutation in Gravity and Cosmology}},
  \href{https://doi.org/10.1142/S0217751X21300064}{\emph{Int. J. Mod. Phys. A}
  {\bfseries 36} (2021) 2130006}
  [\href{https://arxiv.org/abs/2012.11608}{{\ttfamily 2012.11608}}].

\bibitem{Simeon:2020lkd}
G.~Simeon, \emph{{Scalar-tensor extension of Natural Inflation}},
  \href{https://doi.org/10.1088/1475-7516/2020/07/028}{\emph{JCAP} {\bfseries
  07} (2020) 028} [\href{https://arxiv.org/abs/2002.07625}{{\ttfamily
  2002.07625}}].

\bibitem{Reyimuaji:2020bkm}
Y.~Reyimuaji and X.~Zhang, \emph{{Warm-assisted natural inflation}},
  \href{https://doi.org/10.1088/1475-7516/2021/04/077}{\emph{JCAP} {\bfseries
  04} (2021) 077} [\href{https://arxiv.org/abs/2012.07329}{{\ttfamily
  2012.07329}}].

\bibitem{Reyimuaji:2020goi}
Y.~Reyimuaji and X.~Zhang, \emph{{Natural inflation with a nonminimal coupling
  to gravity}},
  \href{https://doi.org/10.1088/1475-7516/2021/03/059}{\emph{JCAP} {\bfseries
  03} (2021) 059} [\href{https://arxiv.org/abs/2012.14248}{{\ttfamily
  2012.14248}}].

\bibitem{Zhang:2021ppy}
X.~Zhang, C.-Y.~Chen and Y.~Reyimuaji, \emph{{Modified gravity models for
  inflation: In conformity with observations}},
  \href{https://doi.org/10.1103/PhysRevD.105.043514}{\emph{Phys. Rev. D}
  {\bfseries 105} (2022) 043514}
  [\href{https://arxiv.org/abs/2108.07546}{{\ttfamily 2108.07546}}].

\bibitem{Chen:2022dyq}
C.-Y.~Chen, Y.~Reyimuaji and X.~Zhang, \emph{{Slow-roll inflation in f(R,T)
  gravity with a RT mixing term}},
  \href{https://doi.org/10.1016/j.dark.2022.101130}{\emph{Phys. Dark Univ.}
  {\bfseries 38} (2022) 101130}
  [\href{https://arxiv.org/abs/2203.15035}{{\ttfamily 2203.15035}}].

\bibitem{Salvio:2021lka}
A.~Salvio, \emph{{Natural-scalaron inflation}},
  \href{https://doi.org/10.1088/1475-7516/2021/10/011}{\emph{JCAP} {\bfseries
  10} (2021) 011} [\href{https://arxiv.org/abs/2107.03389}{{\ttfamily
  2107.03389}}].

\end{thebibliography}\endgroup

\end{document}